\documentclass[1p,onecolumn,authoryear,preprint,times,10pt]{elsarticle-ed}

\journal{arXiv}

\usepackage[english]{babel}
\usepackage[T1]{fontenc}
\usepackage[utf8]{inputenc}
\usepackage{amsmath,amssymb}
\usepackage{booktabs}
\usepackage[dvipsnames]{xcolor}
\usepackage{soul}
\usepackage{fullpage}
\usepackage{lineno}   
\usepackage{subfig}   
\usepackage{latexsym} 
\usepackage{bm}       

\setstcolor{red}
\sethlcolor{yellow}
\setulcolor{green}

\def\citeapos#1{\citeauthor{#1}'s (\citeyear{#1})}

\begin{document}



\begin{frontmatter}

\title{\textbf{\LARGE Shadowing the rotating annulus. Part I:\\Measuring candidate trajectory shadowing times}}

\author[lse,aopp]{\textbf{Roland~M.~B.~Young}\corref{cor1}}
\author[lse]{\textbf{Roman~Binter}}
\author[lse,bios]{\textbf{Falk~Nieh\"{o}rster}}

\cortext[cor1]{Corresponding author. Current address: College of Science, UAE University, P.O. Box 15551, Al Ain, United Arab Emirates. \textit{Email address:} roland.young@uaeu.ac.ae.}

\address[lse]{Centre for the Analysis of Time Series, London School of Economics, London, UK}
\address[aopp]{Atmospheric, Oceanic and Planetary Physics, Department of Physics, University of Oxford, Oxford, UK}
\address[bios]{Bermuda Institute of Ocean Sciences, St. George's, Bermuda}

\begin{abstract}
An intuitively necessary requirement of models used to provide forecasts of a system's future is the existence of {\it shadowing} trajectories that are consistent with past observations of the system: given a system-model pair, do model trajectories exist that stay reasonably close to a sequence of observations of the system? Techniques for finding such trajectories are well-understood in low-dimensional systems, but there is significant interest in their application to high-dimensional weather and climate models. We build on work by Smith et al. [2010, {\it Phys. Lett. A}, 374, 2618--2623] and develop a method for measuring the time that individual ``candidate'' trajectories of high-dimensional models shadow observations, using a model of the thermally-driven rotating annulus in the perfect model scenario. Models of the annulus are intermediate in complexity between low-dimensional systems and global atmospheric models. We demonstrate our method by measuring shadowing times against artificially-generated observations for candidate trajectories beginning a fixed distance from truth in one of the annulus' chaotic flow regimes. The distribution of candidate shadowing times we calculated using our method corresponds closely to (1) the range of times over which the trajectories visually diverge from the observations and (2) the divergence time using a simple metric based on the distance between model trajectory and observations. An empirical relationship between the expected candidate shadowing times and the initial distance from truth confirms that the method behaves reasonably as parameters are varied.\\

\noindent \textbf{This paper was submitted to \textit{Physica D} in 2010, but, after review, was not accepted. We no longer have the time or resources to work on this topic, but would like this record of our work to be available for others to read, cite, and follow up.\\}

\end{abstract}

\begin{keyword}
Shadowing trajectories; Rotating annulus; Dimensional scaling; Weather and climate forecasting; Chaotic dynamics; Perfect Model Scenario.
\end{keyword}

\end{frontmatter}

\section{Introduction}

In forecasting one is often forced to use complex numerical models to predict the future behaviour of physical or mathematical systems. A good example are General Circulation Models (GCMs), which are used to predict the future behaviour of the atmosphere. GCMs have state spaces of over ten million dimensions and integrate differential equations describing complex dynamical and chaotic behaviour. Decisions on how to act over future timescales of days (in the case of weather) up to decades (in the case of climate) are made based on the output from these models, and so decision makers desire these forecasts to be informative.

An intuitive requirement of models used to provide forecasts of a system's future is the existence of model trajectories that are consistent with past observations of the system. Given a system-model pair, do model trajectories exist that stay reasonably close to a sequence of observations of the system? Such trajectories are said to \textit{shadow} the observations. While the existence of a shadowing trajectory is not a \textit{sufficient} condition for successful forecasts, it is intuitively a \textit{necessary} one. 

The concept of shadowing appears in several slightly different contexts, and it is important to differentiate between these. In its original context, so-called $\epsilon$-shadowing, the shadowing problem is concerned with whether there exists a true solution to a differential equation that remains sufficiently close to a given numerical solution. If no true solution exists that shadows the numerical solution, can we learn anything at all about the true solution from the numerical solution? The $\epsilon$-shadowing time is formalised in the Anosov-Bowen shadowing lemma \citep{1975Bowen}. Second, a superficially similar but fundamentally different context is $\iota$-shadowing \citep{1998Gilmour}. This is concerned with whether a trajectory of a numerical model exists that shadows a given sequence of observations of a system. The $\iota$-shadowing time is the maximum model trajectory length that, among all possible trajectories, remains consistent with the observational noise, and is defined with respect to a particular noise model \citep[p.~47]{1998Gilmour}. In this context, if the model does not admit such a trajectory then there is no such initial condition that remains close to the observations, and the problem is then one of model error, not chaos. Third and finally, the $\phi$-shadowing problem \citep[p.~52]{2000SmithA} is concerned with whether there exists a model trajectory that remains close enough to a given set of observations for ``useful'' forecasts to be possible. As what is ``useful'' depends on context, this use of the term is necessarily subjective. For example, one could argue that a GCM solution that predicts the wrong amount of rain in a given place but does predict that rain will fall is ``useful'' enough to constitute a $\phi$-shadowing trajectory.

In the atmospheric context we are interested in how well particular models shadow observations. Several methods exist that attempt to find shadowing trajectories directly from a sequence of observations \citep[e.g.][]{1998Gilmour,2009Stemler}, and these methods are well-understood using low-dimensional (low-$N$) systems such as the Lorenz equations \citep{1963Lorenz} ($N=3$) and the Ikeda map \citep{1979Ikeda} ($N=2$). Whether these methods are feasible when applied to high-dimensional models of the atmosphere remains poorly understood, however, as is the time over which atmospheric models can shadow past observations.

\begin{figure}[tb]
  \centering
  \includegraphics[viewport = 175 80 420 330, clip,width=0.45\textwidth]{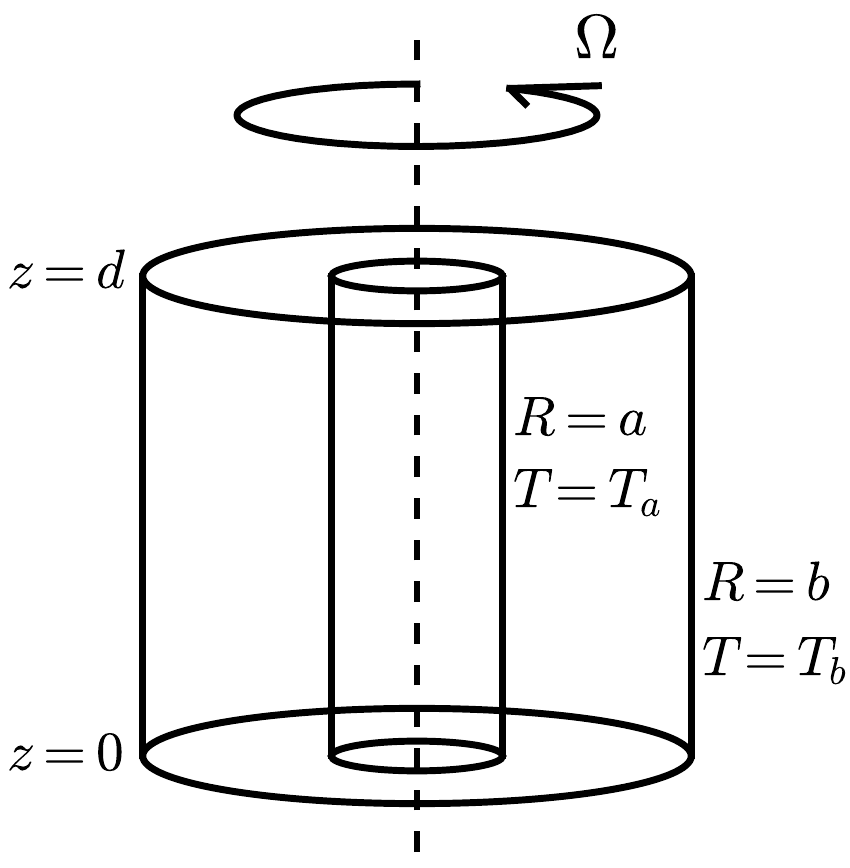}
  \caption{\small Schematic of the thermally-driven rotating annulus experiment. The inner cylinder is at radius $a=2.5$\,cm, the outer cylinder is at radius $b=8.0$\,cm, and the cylinders have a depth of $d=14.0$\,cm. The fluid between the cylinders is a mixture of glycerol and water, 17\%/83\% by volume. The cylinders are maintained at constant temperatures $T_a=18$\,$^{\circ}$C and $T_b=22$\,$^{\circ}$C respectively, giving a constant temperature difference across the fluid of $\Delta T=4$\,degC. The whole system is rotated about a vertical axis with angular velocity $\Omega=1.00$\,rad\,s$^{-1}$.}
  \label{fig:annulus-schem}
\end{figure}

Intermediate in complexity between low-dimensional systems and the atmosphere are laboratory experiments, such as the rotating annulus \citep{1953Hide,1975Hide,1985HignettB,1992ReadA}. This system consists of a fluid contained between two concentric cylinders, rotating about a vertical axis with a horizontal temperature gradient between the cylinders (Fig.~\ref{fig:annulus-schem}). This setup mimics the major influences acting on a planet's atmosphere: the effects of rotation, gravity, and the temperature difference between low and high latitudes. The rotating annulus has been well-established over 50 years as a useful analogue for certain types of atmospheric dynamical behaviour, especially the general circulation of the atmosphere at mid-latitudes \citep{1961Winston,1967Lorenz}. Models used to simulate the behaviour of fluid in the annulus are intermediate in complexity between low-dimensional systems and global atmospheric models; the dimension of these models are $O(10^4)$, compared with $O(10)$ for low-dimensional systems and $O(10^7)$ for GCMs. Experimental systems are also intermediate in their physical idealization of the system being studied, between GCMs with a low degree of idealization, and systems such as the Lorenz equations, which describe a highly-idealized representation of convective rolls at the onset of convection. This paper is the first in a larger body of work exploring shadowing techniques using the rotating annulus. We aim to understand shadowing methods better by examining them in a real but idealised system, and also improve our understanding of the annulus and the models used to study it through their shadowing properties.

Later work will address the problem of finding the longest model trajectories that shadow a given sequence of observations. Here we focus on how to measure the shadowing time itself. Most methods that look for shadowing trajectories\footnote{\citet{1988Hammel} is an exception, but would be very difficult in practice to extend to a high-dimensional system.} generate a set of \textit{candidate} trajectories, each of which may or may not shadow the observations. One then has to measure how long each candidate shadows, and the longest of these \textit{candidate shadowing times} gives a lower bound on the true shadowing time for that model --- observation pair. The main aim of this paper is to present a method for identifying shadowing times that can be used with high-dimensional systems, like the rotating annulus, when the observational error distribution is unbounded. We begin from a definition outlined in \citet{2010Smith} which has some problems when applied to high-dimensional systems, and present a modified form that solves these problems. 

We use the annulus to illustrate the method by examining some of the shadowing properties of our annulus model using the perfect model scenario (PMS). The aim is to demonstrate that our method yields shadowing times that are reasonable. For example, we build a statistical distribution showing how long a candidate might expect to shadow observations if it starts a particular distance from the system's true state. When the initial conditions lie off the system's attractor then a reasonable method for identifying shadowing times in the PMS should, in general, yield longer times the closer the initial condition is to truth. This result will establish a benchmark for future work when we search for shadowing trajectories by generating candidates from observations using the gradient descent method \citep{2003Judd,2004JuddB}. As a sanity check we also use a simple distance-based metric to determine whether our shadowing times are reasonable; they should capture the time the model visually diverges from the observations. The PMS provides a robust framework for investigating shadowing in complex systems, as it allows many aspects of the setup to be controlled, and it also allows us to avoid dynamical problems due to the $\epsilon$-shadowing properties of the model.

In Sect.~\ref{sec:tests} we present an analysis of the \citet{2010Smith} shadowing definition and describe our alternative for high-dimensional systems. Then in Sect.~\ref{sec:method} we describe the rotating annulus model and the experimental setup, before investigating how the shadowing time depends on the distance of the candidate state from truth in Sect.~\ref{sec:results}. Finally we discuss our results and draw conclusions in Sect.~\ref{sec:conc}.

\section{A new method for measuring shadowing times in high-dimensional systems}
\label{sec:tests}

\subsection{The perfect model scenario}
\label{sec:pms}

Consider a model $f$ of a system $\tilde{f}$. If the model is a perfect representation of the system, $f\equiv\tilde{f}$, then there always exists a trajectory of the model that produces an infinitely long perfect shadow of the system. To investigate the limits of shadowing it is useful to work in the perfect model scenario (PMS), where the model and the system are equivalent ($f\equiv\tilde{f}$), but noise does obscure the system state. In the PMS, the only limit to shadowing the system is the uncertainty in a trajectory's initial state. When modelling physical systems this is never the case, because models are always imperfect representations of the system ($f\not\equiv\tilde{f}$). We call this setting the imperfect model scenario (IMS), and then model imperfections and dynamical aspects of shadowing a numerical solution (see below) limit the shadowing capabilities of the model. In practice, in real systems model imperfections will be the overwhelmingly larger of the two.

In this paper we only use the PMS; the main advantage being that we know the true state of the system. This allows us to generate observations by purposely obscuring the true state by a known noise model of observational error ${\bf \Sigma}$. Controlled experiments can then provide insights into the properties of shadowing trajectories under initial condition uncertainty alone. We measure how long a model trajectory started a given distance from the system state $f({\bf x}+{\bf \delta})$ stays consistent with observations of the system $\tilde{f}({\bf x})+{\bf e}$, where ${\bf e}\sim{\bf \Sigma}$ represents a realisation of the observational noise model. In the future, our aim is to explore the IMS as well by allowing the model to be imperfect; \citet{2002SmithA} discusses the subtleties of the PMS and IMS in more detail.

There is a rich literature on the shadowing of numerical solutions of differential equations by their true solutions, the so-called $\epsilon$-shadowing problem \citep[for example]{1987Hammel,1990Grebogi,1991SauerB,1997Sauer,1999Lai}. Representing differential equations on a computer introduces round-off error into the numerical solution at each timestep, which tends to grow exponentially when the system is chaotic until the analytical and numerical solutions diverge \citep{1987Hammel}. In the current context the PMS guarantees that this problem does not arise. By construction it is the numerical solution itself that is the system $\tilde{f}$, \textit{not} the underlying equations. Hence the map from time $t$ to $t+\delta t$ is the same for both model $f$ and system $\tilde{f}$, \textit{both} including the same numerics for a given $\bf x$, and so by construction we avoid the $\epsilon$-shadowing problem entirely. Even when shadowing the laboratory experiment this phenomenon will not be relevant, because  model imperfections will be the dominant source of error, leading to model drift \citep[see, e.g.,][for an example in weather modelling]{2001Orrell}.

\subsection{Consistency in shadowing}

The \textit{candidate shadowing time} $\tau_S$ for a specific candidate trajectory is the time that trajectory remains consistent with observations, given a model of observational error. \citet{2010Smith} define the two as consistent if the distribution of the forecast residuals, ${\bf e_t\sim{\bf \Phi}}$, and the distribution of noise realisations given by the noise model, ${\bf \tilde{e}}\sim{\bf \Sigma}$, are equivalent. As long as ${\bf \Phi}\equiv{\bf \Sigma}$, the candidate trajectory is said to shadow the observations. For a particular model and observations, the \textit{shadowing time} for that pair is the maximum among all possible candidates: $\tau_{S_t}=\max_{\bf x}\tau_S({\bf x},t)$. By measuring $\tau_S$ for a finite set of candidate trajectories one can only calculate a lower bound on $\tau_{S_t}$. 

In practice ${\bf \Phi}$ is expected to differ from ${\bf \Sigma}$ and we aim to quantify how similar the two distributions are. The degree of similarity between ${\bf \Phi}$ and ${\bf \Sigma}$ can be assessed by comparing equivalent percentiles of the two distributions. Suppose we have a single sample $x_1,\ldots,x_N$ from a distribution ${\bf \Phi}$ and want to determine whether 
$\bf\Phi$ is close to some known distribution $\bf\Sigma$. We can proceed by computing the $r$-th order statistic $X_{(r)}$ of the sample from $\bf\Phi$ and perform a statistical test to determine whether $X_{(r)}$ is close to the $r$-th order statistic $\tilde{X}_{(r)}$ from a sample $\tilde{x}_1,\ldots,\tilde{x}_N$ from $\bf\Sigma$. This calculation requires the distribution $\tilde{F}_{(r)}$ of the order statistic $\tilde{X}_{(r)}$. The intuitive way to do this is to repeatedly sample $N$ values $\tilde{x}_1,\ldots,\tilde{x}_N$, from $\bf\Sigma$, calculate $\tilde{X}_{(r)}$ for each sample, and construct the empirical distribution $\tilde{F}_{(r)}$. We can then check whether $X_{(r)}$ falls within a particular confidence interval of $\tilde{X}_{(r)}$. If it does, we say that the order statistic $X_{(r)}$ is consistent with the distribution $\tilde{F}_{(r)}$ of the order statistics $\tilde{X}_{(r)}$ and hence the two distributions are sufficiently close.

\begin{figure}[tb]
  \centering
  \includegraphics[width=0.5\textwidth,clip,viewport=100 360 762 700]{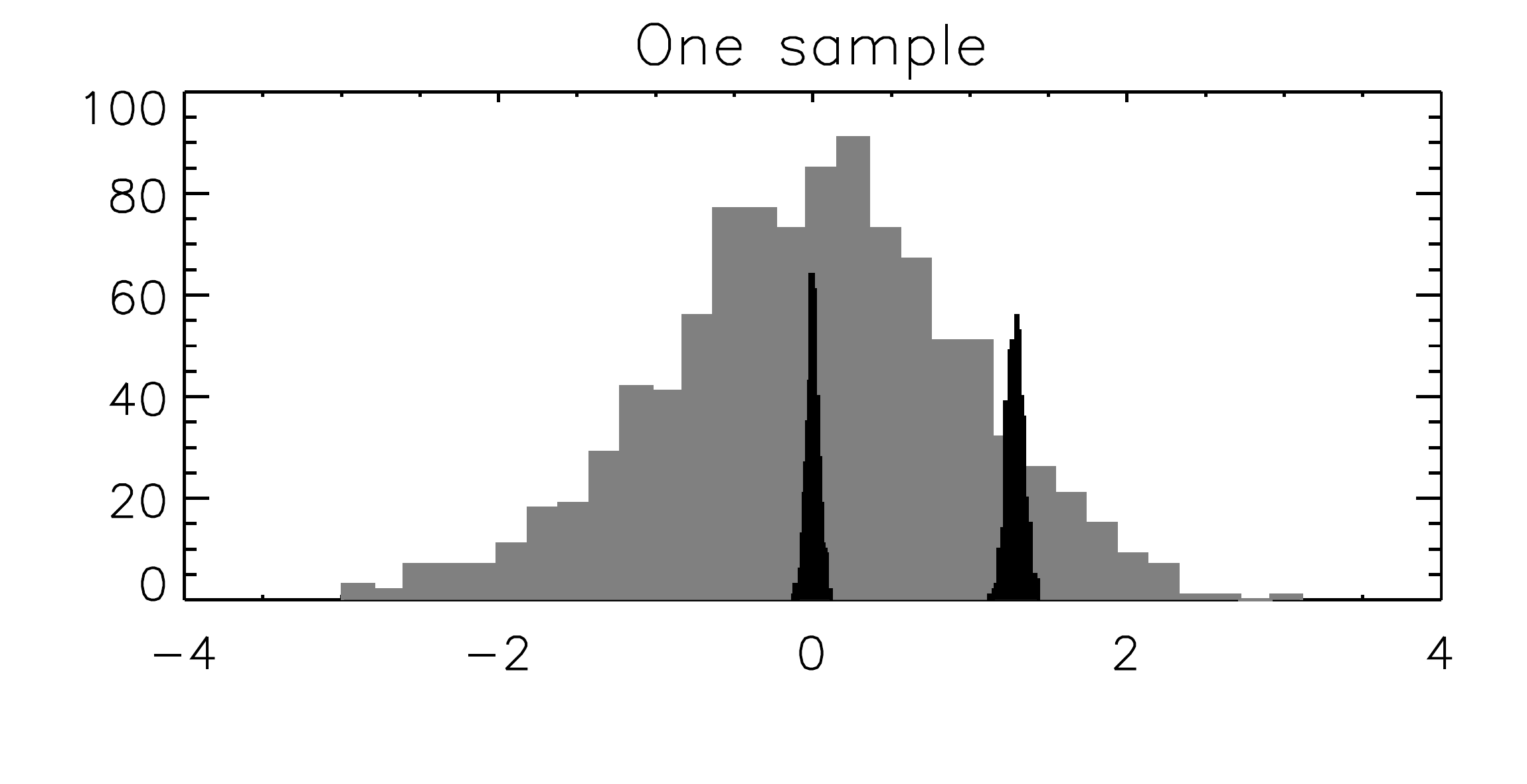}\\
  \includegraphics[width=0.5\textwidth,clip,viewport=100 360 762 700]{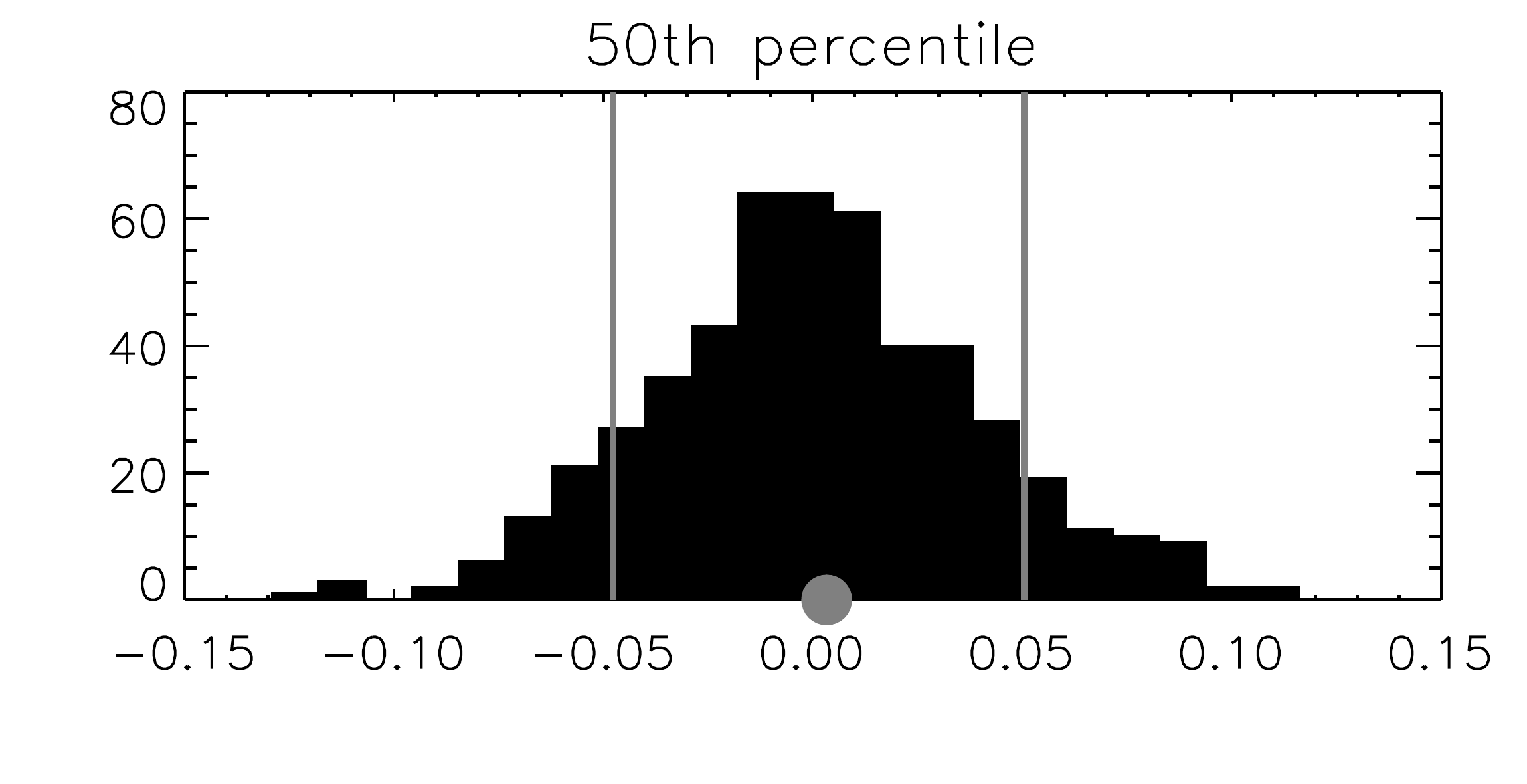}\\
  \includegraphics[width=0.5\textwidth,clip,viewport=100 360 762 700]{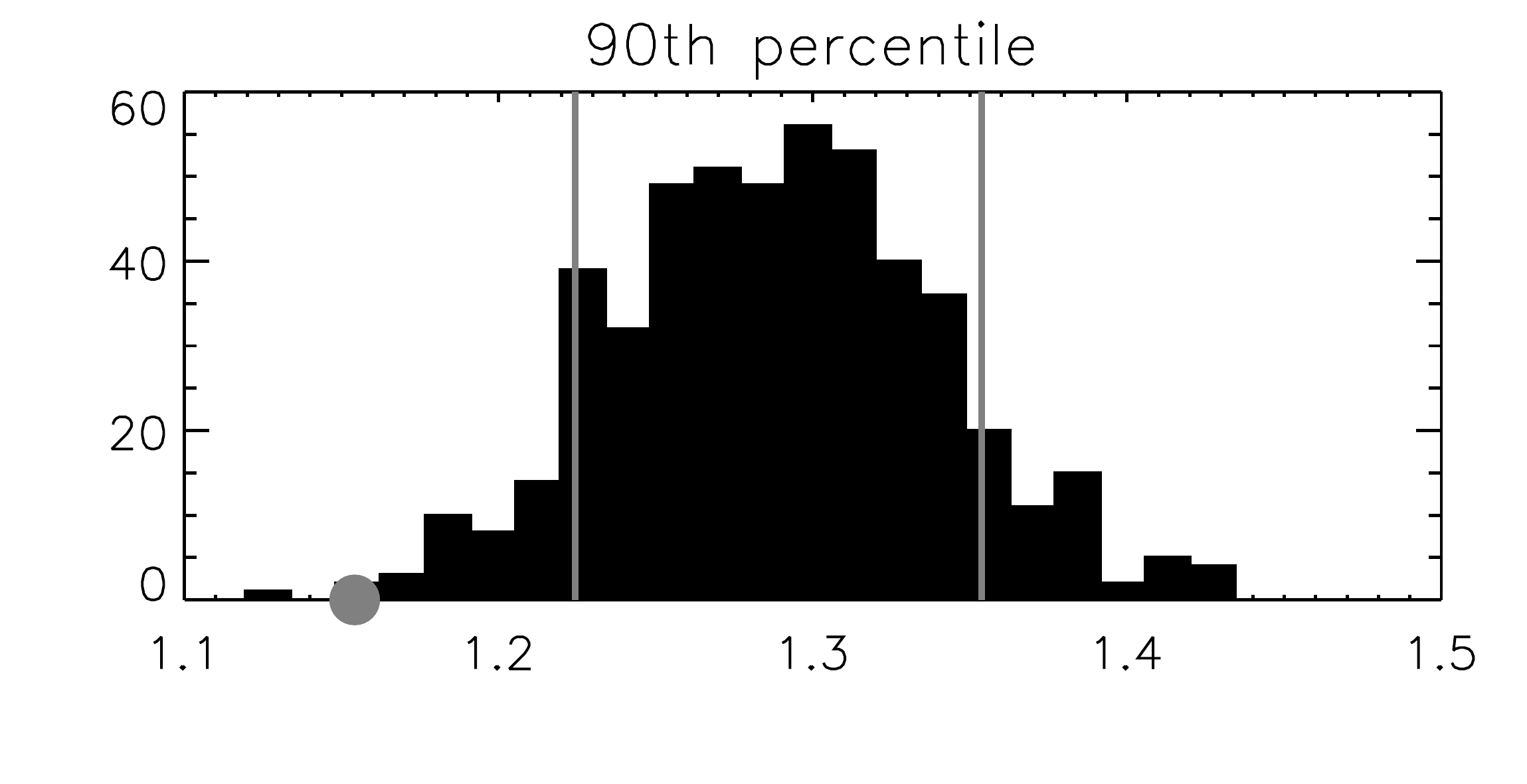}
  \caption{\small Schematic of the general order statistics approach to calculating a candidate's shadowing time. The top panel shows a histogram (grey) of the values generated from a sample of size 1000 from $N(0,1)$. The black histograms show, for 500 such samples, the distributions of the 50th and 90th percentiles of each sample. These percentile distributions are shown in the middle (50th) and lower (90th) panels. On the lower panels vertical lines are added at percentiles 10\% and 90\% in those distributions. The 50th and 90th percentiles of a single test distribution drawn from $N(0,1.01)$ are shown as grey circles.}
  \label{fig:orderstat-schem}
\end{figure}

For a more robust assessment more than one order statistic $X_{(r)}$ can be tested in this way, which we illustrate in Fig.~\ref{fig:orderstat-schem}. In that figure we draw 500 samples of size 1000 from a known noise distribution ${\bf\Sigma}\sim N(0,1)$, and calculate the 50th ($\tilde{X}_{(50\%)}$) and 90th ($\tilde{X}_{(90\%)}$) percentiles of each sample. The black bars in the figure show the distributions of these percentiles. We then produce a test sample of the same size but from a slightly different distribution ${\bf\Phi}\sim N(0,1.01)$, and calculate its order statistics $X_{(50\%)}$ and $X_{(90\%)}$, shown by circles in the lower figures. To retain the null hypothesis that the test sample is drawn from the same distribution as the noise model at significance level $p=0.2$, the two circles must fall within the vertical lines enclosing the 80\% confidence intervals of the percentile distributions. In this particular example they do not, so the two distributions are not sufficiently close.
  
Applying this procedure to a sample of residuals ${\bf e}_t$ and realizations of a noise model ${\bf 
\tilde{e}}_t$ determines whether the model trajectory is sufficiently close to shadow the target at a given significance level.    

\subsection{The \citet{2010Smith} method for measuring shadowing times}

\citet{2010Smith} use the the Euclidean length of the residual vector $\Vert{\bf e}_t\Vert$ as the test statistic to measure the shadowing time in what we shall refer to as the ``sequence'' method. This is the distance in the $N$-dimensional model state space between the model trajectory and the target. They calculate this quantity at each time and form a distribution of distances over a trajectory time series, before calculating the median and 90th percentile of that distribution.  

To obtain a sampling distribution of a given order statistic of the noise model, they draw a sample of noise realizations ${\bf \tilde{e}}$ of size $N$ from $\bf\Sigma$ at each time. They calculate the Euclidean distance $\Vert{\bf \tilde{e}}\Vert$ for each time, giving a distribution of noise distances, and then locate the median and 90th percentile of \textit{this} distribution. Repeating this sampling procedure $M$ times yields a distribution of $M$ medians and $M$ 90th percentiles under the noise model. To determine whether the model trajectory shadows the target at this time, they test the following null hypothesis: for \textit{both} order statistics (the 50th and 90th percentiles) the percentile of the residual distribution is drawn from the distribution of samples from the noise distribution. The comparison of medians tests the residual distribution for bias, and the comparison of the 90th percentiles tests it for dispersion. They use confidence intervals of 90\% for the median and 99\% for the 90th percentile.  The procedure begins by using the trajectory segment with $t=1$ only. If this segment shadows the target then they repeat with the trajectory segment increased in length by one, and so on, increasing the trajectory segment by one point at a time. The candidate shadowing time is determined once a trajectory segment rejects the null hypothesis.

\subsubsection{The need for an alternative method}
\label{sec:problems}

We see three problems with the \citet{2010Smith} definition that are particularly important when $N$ becomes large.

First, \citeapos{2010Smith} sampling approach is computationally expensive. At each time one requires $M$ samples of size $N$ from the noise model to produce the distribution of percentiles. Of course $M$ should be as large as possible to ensure the best correspondence to the true underlying distribution of each percentile. For systems of very low dimensionality this is feasible but in higher dimensions this sampling approach is no longer computationally practical.    

Second, a single outlier at one point in the model state space in a very short trajectory may distort the distribution of the $\Vert{\bf e_t}\Vert$ and hence cause the model to fail to shadow for that trajectory length, while a longer trajectory including that point would reduce the effect of that outlier in the distribution and hence that longer time series might then shadow. 

Finally, the length of the trajectory is important for another reason when using a distance method. If the trajectory is only length five, for example, then the distribution of distances is too small to obtain a sensible value of, say, the 90th percentile of that distribution. For a sensible shadowing time, therefore, the trajectory must be sufficiently long for statistical accuracy, but often the model is not capable of shadowing for the minimum of, say, $O(100)$ leadtimes required.

\subsection{The ``state'' method for measuring shadowing times}

To avoid some of these problems, we now present a slightly modified version of the method used to measure the shadowing time from the consistency between model trajectory and observations that is better suited to higher-dimensional systems. We also believe this method is more intuitive than using a distribution of distances from model to target along a trajectory. We call this the ``state'' method. 

\subsubsection{Avoiding sampling problems in high-dimensions}

First we present a theoretical result that overcomes the sampling impracticality of the sequence method when applied to cases with $\gg O(10)$ dimensions, opening it up for use in the higher-dimensional context.

Following \citet{1981David}, assume independently and identically distributed (IID) samples of size $N$ from a cumulative distribution function (CDF) $P(x)$, each sample providing $N$ random variables $X_1,\ldots,X_N$. When the sample elements are sorted in ascending order, i.e. $X_{(1)}\leq X_{(2)}\ldots\leq X_{(N)}$, we call $X_{(r)}$ the $r$th order statistic. Denote by $F_{(r)}$ the CDF of the $r$th order statistic $X_{(r)}$. \citet[pp.~8--9]{1981David} showed that $F_{(r)}$ is
\begin{linenomath*}\begin{eqnarray}
F_{(r)}(x) &=& Pr\lbrace X_{(r)}\leq x\rbrace\label{eq:F}\\
&=& \sum^{N}_{i=r} {N\choose i}P^i(x)\left[1-P(x)\right]^{N-i}\label{eq:F_binomial}
\end{eqnarray}\end{linenomath*}
The relation between the binomial sum and the incomplete $\beta$-function~\citep[Eq.~8.17.5]{2010NIST} means we can replace Eq.~(\ref{eq:F_binomial}) with
\begin{linenomath*}\begin{equation}
F_{(r)}(x) = I_{P(x)}(r,N-r+1)\label{eq:relationFI}
\end{equation}\end{linenomath*}
Suppose we are interested in the 90th percentile of the distribution of the $r$th order statistic 
$X_{(r)}$ (i.e. $F_{(r)}=0.90$), with $X$ being sampled from $P(x)$. We require the $x$ that corresponds to the 90th 
percentile of the distribution of $X_{(r)}$. Since $F_{(r)}=I_{P(x)}(r,N-r+1)$, we may find the probability that $P(x)$ assigns the required value $x$ by taking the inverse of the incomplete beta function, i.e. 
\begin{linenomath*}\begin{equation}
  P(x) = I^{-1}_{F_{(r)}}(r,n-r+1)
\end{equation}\end{linenomath*}
Finally we obtain the value $x$ which gives the required percentile of the distribution of $X_{(r)}$ using the inverse of $P(x)$: $x=P^{-1}(x)$.       

This theoretical result provides an exact distribution of any order statistic $X_{(r)}$ at negligible computational cost. We have overcome the intractability of sampling in high dimensions, and so under the assumption of IID samples we can now perform formal hypothesis testing on a sample of any size.

\subsubsection{Definition and properties of the ``state'' method}

With this result in mind, we present a modified form of the \citet{2010Smith} sequence method used to calculate shadowing times. Our method overcomes the problems listed in Sect.~\ref{sec:problems} above while remaining computationally practical. 

At each time $t$ in a model trajectory we have a model state ${\bf x}_{t}$, a target state ${\bf \tilde{x}}_{t}$ (whose observational error is well-defined by the distribution $\bf\Sigma$ in the PMS), and a residual error vector ${\bf e}_{t}={\bf x}_{t}-{\bf\tilde{x}}_{t}$, each with dimension $N$. Starting at $t=0$, we test the null hypothesis at significance level $p$ that the residual error vector ${\bf e}_{t}$ is a sample drawn from the observational error distribution $\bf\Sigma$. To perform the test we measure the median and the 90th percentile of the residual error vector, and find the distributions of the equivalent percentiles of the noise distribution analytically using the order statistics method discussed above. To shadow the observations at time $t$ we require both the median and 90th percentiles of the residual distribution to fall within the respective confidence intervals of the noise distribution at significance level $p$. If the model shadows at $t=0$ then we repeat the procedure for $t=1,2,\ldots$, and the shadowing time for that model trajectory is the longest time for which the null hypothesis is retained.

This method differs from the ``sequence'' method in a number of ways. First, we avoid sampling the noise distribution by using the inverse $\beta$-function as described in the previous section\footnote{Note that now this result has been obtained, sampling can be avoided for the ``sequence'' method as well.}. This avoids problem one in Sect.~\ref{sec:problems}. Second, our method applies to each point in the model trajectory in turn, rather than using the whole trajectory at once. Our method uses the distribution of values over the components of the state vector, while \citet{2010Smith} map the vector of components to a single number. This property means we avoid problem two in Sect.~\ref{sec:problems}. Finally, we avoid problem three in Sect.~\ref{sec:problems} by requiring the state vector dimension $N$ to be large. Therefore our method is not appropriate for use with low-dimensional systems, where $N$ is not large enough to compute sensible percentile statistics for the residual error distribution. Take, for instance, the \citet{1963Lorenz} system, with $N=3$. Determining the 90th percentile in a sample of three values does not produce a sensible result. Our method is suitable for medium- and high-dimensional systems, with a dimensionality of at least $N=O(100)$ and preferably $N=O(\gg 100)$.

We should note that a caveat with our method is that by examining each trajectory point in turn we may not detect small but systematic outliers. If an outlier is small but present at each trajectory point the model should shadow at that point. If the whole trajectory is considered, however, as in~\citet{2010Smith}, the outliers at all times will accumulate and the failure of the model trajectory will be detected.           

\section{Experimental methods}
\label{sec:method}

To demonstrate our method for measuring shadowing times from the previous section and investigate some shadowing properties of our rotating annulus model we run some perturbation experiments; this section describes how they are set up.

\subsection{The rotating annulus simulation}
\label{sec:annulus}
\label{sec:model}

Figure~\ref{fig:annulus-schem} shows a schematic of the rotating annulus with the particular configuration used in this work, which is the ``small annulus'' setup used by \citet{1985HignettB}. We work in the PMS, so ignore details of the laboratory setup that introduce complications due to imperfections in the apparatus, for example, and we take the numerical representation of Fig.~\ref{fig:annulus-schem} as a perfect representation of the system. 

The mathematical model used to simulate the rotating annulus experiment is the Met Office / Oxford Rotating Annulus Laboratory Simulation (MORALS) \citep{1976Farnell,1985HignettB,2000Read}. This model solves the Navier-Stokes, heat transfer, and continuity equations subject to the Boussinesq approximation, in cylindrical polar coordinates (see Appendix). There are four prognostic variables: three velocity directions $\bf u$ (radial), $\bf v$ (azimuthal), $\bf w$ (vertical), and temperature $\bf T$. The fifth field is a kinetic pressure field ${\bm\Pi}\equiv {\bf p}/\rho_0$, which is diagnostic (calculated using a Poisson equation from the other four fields); $\rho_0$ is the fluid density at a reference temperature. The fields are discretized on a staggered Arakawa C grid \citep{1977Arakawa}, and are non-uniform in the radial and vertical directions to resolve the boundary layers. All velocities are set to zero at the boundaries, and the temperature gradient is zero across the top and bottom boundaries. The parametrisation of fluid properties used to set up a 17\% glycerol / 83\% water fluid (by volume) is described in \citet[][Table~1]{1985HignettB}. The model runs with a timestep of $\delta t=0.02$\,s and a spatial resolution of $(N_R,N_{\theta},N_z)=(16,32,16)$ nodes in the radial, azimuthal, and vertical directions respectively. The number of independent variables\footnote{From $16\times 32\times 16\times 4$ minus the number of points on and outside the fluid boundary (some outside points are required by the boundary conditions when using a staggered grid).} is $N=24192$.

\begin{figure}[tb]
  \centering
  \subfloat[Snapshot of an equilibrated wavenumber-3 temperature structure in a horizontal section at $z=5.61$\,cm after 3000\,s.]{\includegraphics[viewport=20 10 610 540,clip,width=0.49\textwidth]{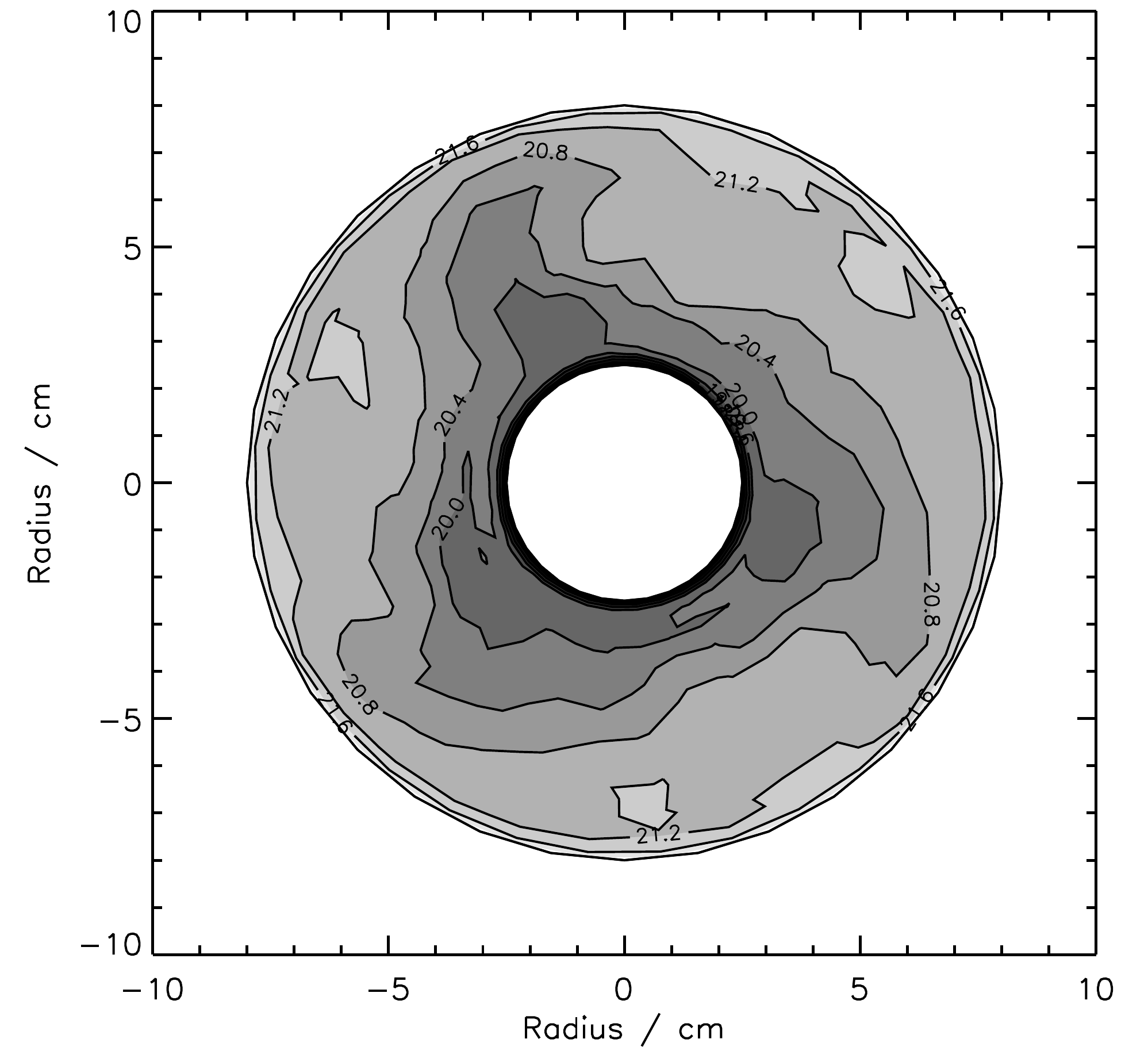}\label{fig:annulus-T-field-slice}}\,\,
  \subfloat[Time series of the wavenumber-3 temperature mode amplitude in the same simulation. The amplitude is found at each timestep from a Fourier transform of the temperature field at mid-radius / mid-height.]{\includegraphics[viewport=0 0 276 190,clip,width=0.49\textwidth]{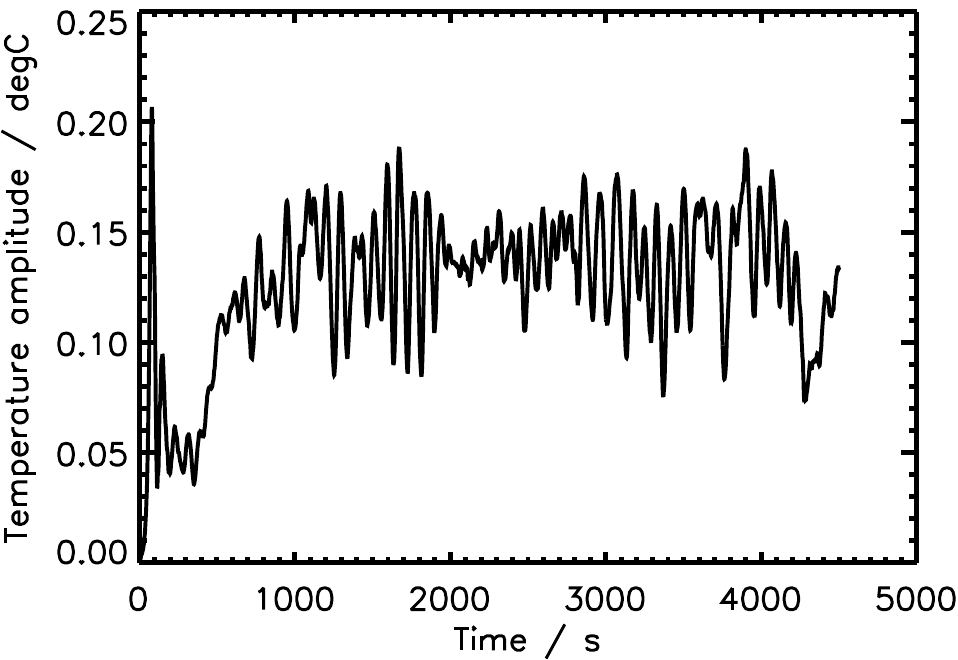}\label{fig:annulus-T-field-tseries}}
  \caption{\small Temperature output from a rotating annulus simulation using the model described in the text with the setup in Fig.~\ref{fig:annulus-schem}.}
  \label{fig:annulus-T-field}
\end{figure}

Fig.~\ref{fig:annulus-T-field} shows temperature output from an annulus simulation using this configuration, with $\Omega=1$\,rad\,s$^{-1}$ and $\Delta T=4$\,degC. Figure~\ref{fig:annulus-T-field-slice} is a typical horizontal temperature field near mid-height, showing a wavenumber-3 structure to the flow, and Fig.~\ref{fig:annulus-T-field-tseries} shows the amplitude of the wavenumber-3 azimuthal temperature mode as a function of time. The flow starts from rest in the rotating frame at $t=0$, so there is some transient behaviour in the first $\sim$1000\,s. 

To determine whether the flow regime is chaotic we estimated the largest Lyapunov exponent for this run. Chaos requires aperiodic behaviour in a deterministic system with at least one positive Lyapunov exponent \citep[p.~323]{1994Strogatz}. To estimate the largest Lyapunov exponent we followed the procedure that \citet{2008YoungA} used on annulus simulations in various flow regimes \citep{1985Wolf,1999Hegger}, in this case working with the wavenumber-3 temperature time series. The procedure was the same as documented in \citet{2008YoungA}, with 62\,s the optimal delay time in this case, and we extended the time series in Fig.~\ref{fig:annulus-T-field-tseries} to 100000\,s to minimise noise in the calculation. We found the largest Lyapunov exponent to be $0.0061\pm 0.0025{\rm\,bits\,\,s^{-1}}$, which corresponds to an error-doubling time of 1.5--3.0 oscillation periods (125--250\,s). This exponent is comparable to values in the chaotic structural vacillation flow regime (0.0013--$0.0055{\rm\,bits\,\,s^{-1}}$), and is an order of magnitude larger than values in non-chaotic flow regimes (-0.00012--$0.00063{\rm\,bits\,\,s^{-1}}$) \citep{2008YoungA}, so we might expect perturbations to grow at a similar rate to structurally vacillating flow. The model is deterministic, and the flow is aperiodic for the length of the time series, so the positive Lyapunov exponent demonstrates (for all practical purposes) that this flow is chaotic\footnote{The reader should note that in the laboratory the flow under this external forcing is usually regular wavenumber-3 or 2, which is also the case when running MORALS at higher resolution \citep[Fig.~3b]{2008YoungA}. Because we consider only the PMS in this work, however, direct correspondence with the laboratory experiment is not so important here.}.

\subsection{MORALS perfect model setting and observations}
\label{sec:pms-morals}

Let ${\bf x}[t]$ and ${\bf \tilde{x}}[t]$ be model and true system states at time $t$, and $f^{\Delta t}$ and $\tilde{f}^{\Delta t}$ be integration of states by the model and system for time $\Delta t$. So ${\bf x}[t+\Delta t]=f^{\Delta t}({\bf x}[t])$ and ${\bf \tilde{x}}[t+\Delta t]=\tilde{f}^{\Delta t}({\bf \tilde{x}}[t])$ describe the model and system exactly. Let ${\bf s}[t] = {\bf \tilde{x}}[t]+\tilde{\bf e}$ be observations of the system at time $t$, and $\Delta{\bf x}$ be perturbations to initial states. In the PMS we simply choose the system being modelled, $\tilde{f}$, to be MORALS with the same parameters as the model: $f\equiv\tilde{f}$.

We generate a true system trajectory and observations of that trajectory as follows. First we run MORALS for 1500\,s to spin up the model from rest until a coherent flow structure is present (e.g. Fig.~\ref{fig:annulus-T-field}) and any transient behaviour has decayed. From this we run MORALS for an additional 1500\,s to obtain an estimate, at each grid point, of the range of values admitted by the model when it is in dynamical equilibrium. This gives the natural variability of the system at each point in space. Let the vector ${\bf r}$ be this range, where each element contains the range of values that encloses 99\% of the values over these 1500\,s at that grid point:
\begin{linenomath*}\begin{equation}
  {\bf r}={\bf \tilde{x}}[t_0-1500s,...,t_0]_{99.5\%}-{\bf \tilde{x}}[t_0-1500s,...,t_0]_{0.5\%}
  \label{eq:r}
\end{equation}\end{linenomath*}
The final state in this run (i.e. after a total of 3000\,s) will be the first state in the true system trajectory. To make things simpler later, we redefine this point as $t=0$. We then obtain a long system trajectory by integrating this state using MORALS for time $T$ to obtain ${\bf \tilde{x}}[t],\,t\in[0,\ldots,T]$. 

We then generate observations ${\bf s}[t]$ every 5\,s along the system trajectory by adding a vector of random numbers $\tilde{\bf e}$ (noise) to the system trajectory. The random numbers are independently and identically distributed (IID) and are originally drawn from the Gaussian distribution $N(0,1)$. These random numbers are converted to observational error by multiplying by a fixed fraction $\sigma$ ($=0.1$ throughout this paper) of the natural variability $\bf r$ at each grid point, i.e.
\begin{linenomath*}\begin{equation}
	\tilde{\bf e}\sim\sigma\,{\bf r}\,N(0,1)
	\label{eq:obs-error}
\end{equation}\end{linenomath*}
The effect of the scaling by $\bf r$ ensures that, statistically, the amount of noise added at each point and to each field is the same fraction of the natural variability. For this reason, much of the analysis to follow will be done in terms of quantities expressed as a fraction of the natural variability, as this non-dimensional metric provides a convenient way to compare and combine the variables. We shall call this state space the ``scaled space'', as opposed to the (dimensional) ``physical space''. For two points in the physical state space, ${\bf x}_1$ and ${\bf x}_2$, their separation $\Delta{\bf x}'$ in the scaled space is
\begin{linenomath*}\begin{equation}
	\Delta{\bf x}'\equiv({\bf x}_2-{\bf x}_1)'=({\bf x}_2-{\bf x}_1)\circ{\bf r}^{-1}\equiv\Delta{\bf x}\circ{\bf r}^{-1}
	\label{eq:scaled-space}
\end{equation}\end{linenomath*}
where $\circ$ denotes the Hadamard (pointwise) product.

\subsection{Setup of the perturbation experiments}
\label{sec:perturb}

To investigate the shadowing properties of the rotating annulus model MORALS in the PMS, we calculate trajectories of the model starting from initial \textit{candidate} states perturbed from the true state at $t=0$. We use these \textit{candidate trajectories} to demonstrate the ``state'' method developed in the previous section, by calculating how long each candidate shadows the observations in Sect.~\ref{sec:pms-morals}. By varying the size of the perturbation we shall determine how the distribution of candidate shadowing times depends on the original distance of the candidate from the true state. 

To generate a candidate state perturbed from the true state ${\bf \tilde{x}}[0]$ we add a random perturbation $\Delta{\bf x}$ as follows. First, we draw a random vector $\bf n$ of length $N$ from the normal distribution $N(0,1)$. Second, we scale the Euclidean length of the vector to be $\delta$, i.e.
\begin{linenomath*}\begin{equation}
	{\bf n}'=\delta\frac{\bf n}{\Vert{\bf n}\Vert}
\end{equation}\end{linenomath*}
Finally, to obtain a perturbation in physical space (i.e. the perturbation that is actually added to generate the candidate state), we multiply ${\bf n}'$ by the natural variability at each grid point, $\bf r$. Hence the perturbation $\Delta{\bf x}$ is
\begin{linenomath*}\begin{equation}
  \Delta {\bf x}=\delta\frac{\bf n}{\Vert{\bf n}\Vert}\circ{\bf r}
\end{equation}\end{linenomath*}
By construction,
\begin{linenomath*}\begin{equation}
  \Vert\Delta{\bf x}\circ{\bf r}^{-1}\Vert=\delta
  \label{eq:delta}
\end{equation}\end{linenomath*}
so $\delta$ is the length of the perturbation in the scaled space. All the perturbations of length $\delta$ lie on a $N$-dimensional hypersphere of radius $\delta$ centred on the truth, and the Gaussian distribution guarantees that the perturbations are distributed uniformly in direction. By repeating this procedure any number of candidates may be generated.
 
This perturbation method is rarely used to efficiently sample initial condition uncertainty; usually methods such as breeding vectors \citep{1993Toth}, singular vectors \citep{1995Buizza}, or even so-called ``nightmare'' vectors \citep{1999SmithA} are used, which all aim to give shorter candidate shadowing times by sampling the fastest-growing directions. Here we just use a simple method to demonstrate our method for calculating shadowing times. We note that the distribution of candidate shadowing times would probably be different were we to use these other methods.

\subsection{Measuring candidate shadowing times}
\label{sec:iota-test}
\label{sec:phi-test}

For each candidate state used to begin a model trajectory we calculate the candidate shadowing time $\tau_S$ with respect to the observations using the ``state'' method described in the previous section. At each time $t$ we use the model $f^t({\bf x}[0])$ and observations ${\bf s}[t]$ to construct the residual error vector ${\bf e}[t]=f^t({\bf x}[0])-{\bf s}[t]$. The observational error $\tilde{\bf e}$ is well-defined by Eq.~(\ref{eq:obs-error}). We determine whether the distribution of scaled forecast residuals ${\bf e}[t]\circ{\bf r}^{-1}$ is consistent with the distribution of scaled observational error $\tilde{\bf e}\circ{\bf r}^{-1}$, testing the null hypothesis at significance level $p$ for the 50th and 90th percentiles. If either test statistic lies outside the confidence interval of the noise distribution then the candidate trajectory does not shadow the observations at time $t$. The candidate shadowing time is the final $t$ at which the test is passed, counting upwards from zero.

The shadowing time should capture the time the model diverges from the observations. Hence as a sanity check on our shadowing times we also use a simple measure of how long the model remains close to observations. The Euclidean norm $\Vert {\bf x}\Vert=({\bf x}'{\bf x})^{1/2}$ of a vector of length $N$ whose components are drawn from the Gaussian distribution $N(0,\sigma^2)$ follows a $\chi$-distribution with $N$ degrees of freedom and scale parameter $\sigma$ \citep[p.~236]{1946Cramer}. For $N\to\infty$ the asymptotic form of the $\chi$-distribution is a normal distribution with mean $\sigma\sqrt{N-1/2}$ and variance $\sigma^2/2$ \citep[p.~544--5]{1994Stuart}. So if a vector of $N$ observations $s_i$, $i=1,\ldots,N$ is constructed by adding IID random numbers with distribution $N(0,\sigma^2)$ to a true state $x_i$, $i=1,\ldots,N$, then the expected Euclidean distance from the truth to the observations in our scaled space is very close to $\sigma\sqrt{N}$ for the value of $N$ we use. 

The Euclidean distance in the scaled space between a model candidate trajectory $f^t({\bf x}[0]))$ and the observations ${\bf s}(t)$ is 
\begin{linenomath*}\begin{equation}
   D(t)=\Vert \{f^t({\bf x}[0])-{\bf s}[t]\}\circ{\bf r}^{-1}\Vert
   \label{eq:distance}
\end{equation}\end{linenomath*}
so we say that the candidate trajectory remains close to the observations as long as
\begin{linenomath*}\begin{equation}
   D(t)\leq m\sigma\sqrt{N}
   \label{eq:phi-criterion}
\end{equation}\end{linenomath*}
is satisfied, i.e. once the distance from model to observations becomes greater than $m\sigma\sqrt{N}$ then the candidate trajectory is no longer close. We call this time $\tau_D$. Based on test data we use $m=2$, as this represents a considerable deviation from the expected distance between observations and truth, but it turns out that $\tau_D$ is not very sensitive to $m$. Now, this is basically a simple RMS error metric, and while that is not an optimal error measure for nonlinear models \citep[see, e.g.,][]{1999McSharry}, the purpose of using this metric is to catch when the trajectory is \textit{not} close to the observations, and the RMS error will be large when that is the case, so it is useful as a secondary measure here.

\section{Results}
\label{sec:results}

We are interested in how long a candidate trajectory might expect to shadow observations if the setup of an experiment is perfect and it starts at a given distance from the truth. This is an important benchmark to measure, because in practice with physical systems the setup is never perfect. A shadowing time measured in the imperfect model scenario or using laboratory data is of limited use if we have nothing to compare it against. By using the PMS we can obtain an upper limit to the shadowing properties of a particular system under this type of initial condition uncertainty. We expect a priori that, for perturbations in random directions, the time a trajectory remains close to observations should decrease as its initial distance from truth increases. 

\subsection{Candidate trajectories started from perfect initial conditions}
\label{sec:null}

First, however, we must calibrate the ``state'' method by selecting an appropriate significance level $p$. To do so we ran a ``null'' candidate trajectory started from perfect initial conditions, i.e.\,with $\delta=0$. We measured the shadowing time against 32 different realisations of the observational noise model. By construction, the real shadowing time for this case is infinity, but an intrinsic property of our ``state'' method (and any other test based on repeated significance tests, including the \citet{2010Smith} ``sequence'' method, is that some fraction of the candidates will fail to shadow by a given time, even if the forecast is perfect, because the method is based on significance testing. This fraction depends critically on the significance level at which the null hypothesis is tested. So by calculating the parameters required to avoid this problem we can determine \textit{a priori} the minimum significance level at which our ``state'' method should be used in practice.
 
 \begin{figure}[tb]
  \centering
  \includegraphics[viewport=2 2 597 427,clip,width=0.7\textwidth]{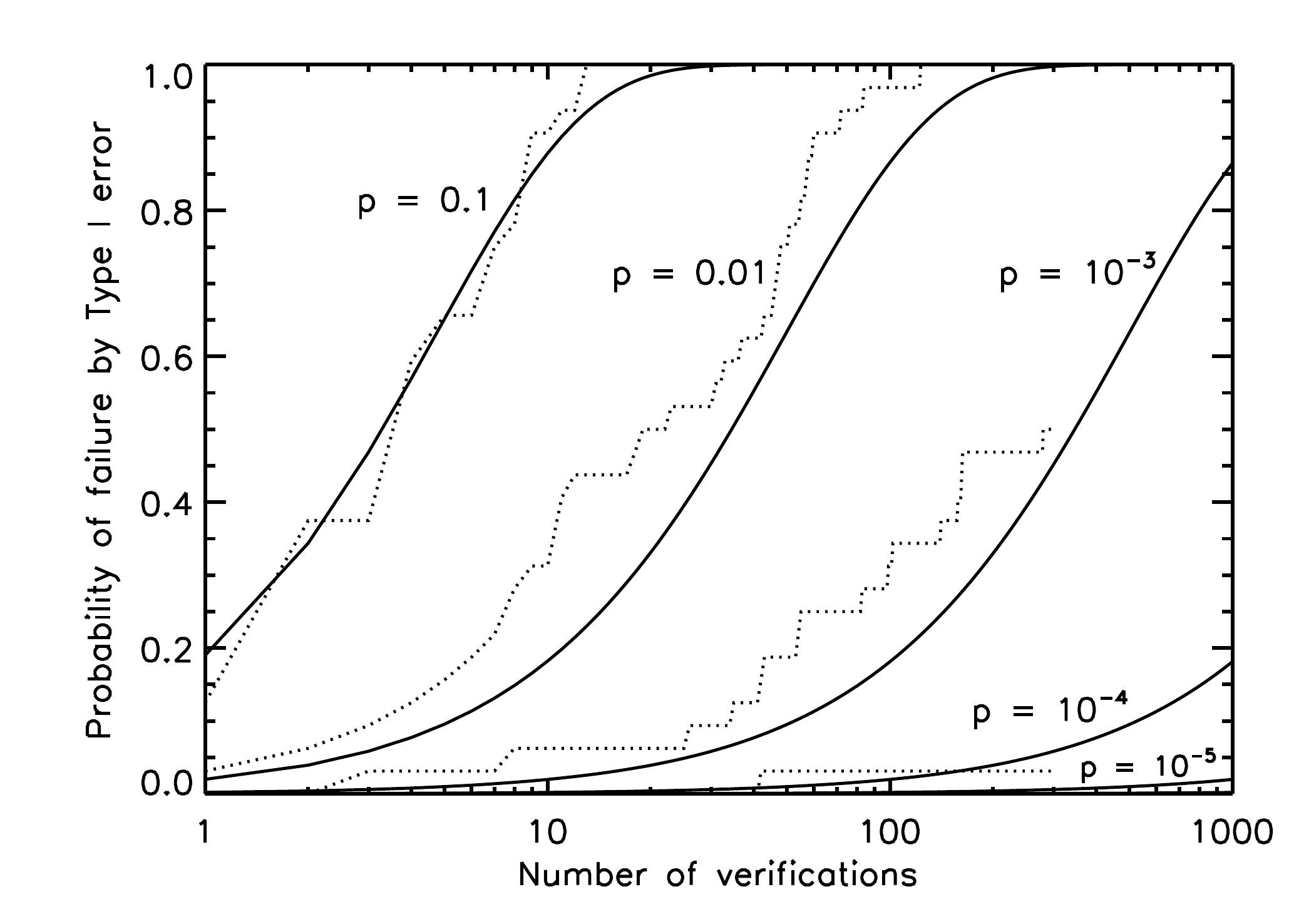}
  \caption{\small Consequences of the Type I error property of the ``state'' method for low significance levels. For each significance level (indicated by its $p$-value) the solid line shows the expected fraction of candidates (from Eq.~\ref{eq:P-type1}) that fail to shadow because of a Type I error as a function of the number of verifications, and the dotted line shows the actual fraction that failed out of the 32 noise realisations. The dotted lines stop at 300 verifications, corresponding to maximum trajectory lengths of 1500\,s.}
  \label{fig:null-test}
\end{figure}

The problem occurs because at each verification time along the trajectory we compute two test statistics, both of which must fall within prescribed confidence intervals in order to retain the null hypothesis at significance level $p$ and hence shadow at that time. The probability of falsely rejecting the null hypothesis when it is true (Type I error) is therefore $1-(1-p)^2$. Within the first $n$ verifications (in this case we check the shadow every 5\,s along the trajectory), the probability of at least one Type I error is
\begin{linenomath*}\begin{equation}
	r = 1-(1-p)^{2n}
  \label{eq:P-type1}
\end{equation}\end{linenomath*}
which is the expected fraction of candidates that will fail to shadow within the first $n$ verifications purely from Type I errors. For significance levels commonly used in statistical testing ($p=0.1$, 0.01, say) this probability becomes large after $O(10)$ verifications. We must avoid these Type I errors because they erroneously assign a candidate shadowing time shorter than the true value.

In Fig.~\ref{fig:null-test} we compare the fraction of candidates that failed to shadow because of a Type I error at each $t$ against theoretical predictions from Eq.~(\ref{eq:P-type1}); the two correspond well. By $t=50$\,s ($n=10$), at significance level $p=0.1$ almost all the candidates have rejected the null hypothesis and hence recorded a (false) shadowing time, as have almost all of the candidates by $t=500$\,s ($n=100$) for $p=0.01$. The 25th percentile of the distribution for $p=0.001$ also falls around this time, there is a sole outlier for $p=0.0001$, and for $p\leq 10^{-5}$ none of the candidates suffer a Type I error.

This property is a fundamental limitation of any method that relies on repeated statistical significance tests, but we can largely avoid it in practice through our choice of $p$. We can find the largest value of $p$ such that in the event that the true shadowing time is the equivalent of $n$ verifications, fewer than $R$ out of a total of $E$ candidates will suffer a Type I error. This condition is $r<R/E$, and hence
\begin{linenomath*}\begin{equation}
	p<1-\left(1-\frac{R}{E}\right)^{1/2n}
	\label{eq:no-type1}
\end{equation}\end{linenomath*}
is required, by rearranging Eq.~(\ref{eq:P-type1}). This problem with Type I errors is well-known in the statistical and medical literature \citep{2007Abdi}; the expression in Eq.~(\ref{eq:no-type1}) is an example of a \textit{\v{S}id\'{a}k equation}.

\subsection{Candidate trajectories started a fixed distance from truth}

Now we demonstrate our method for identifying shadowing times by examining how long candidate trajectories started a fixed distance from truth shadow the observations. 

\begin{figure}[tbp]
  \centering
  \subfloat[Temperature model trajectory (black line) and corresponding observations (grey line).]{\includegraphics[viewport=25 16 597 400,clip,width=0.8\textwidth]{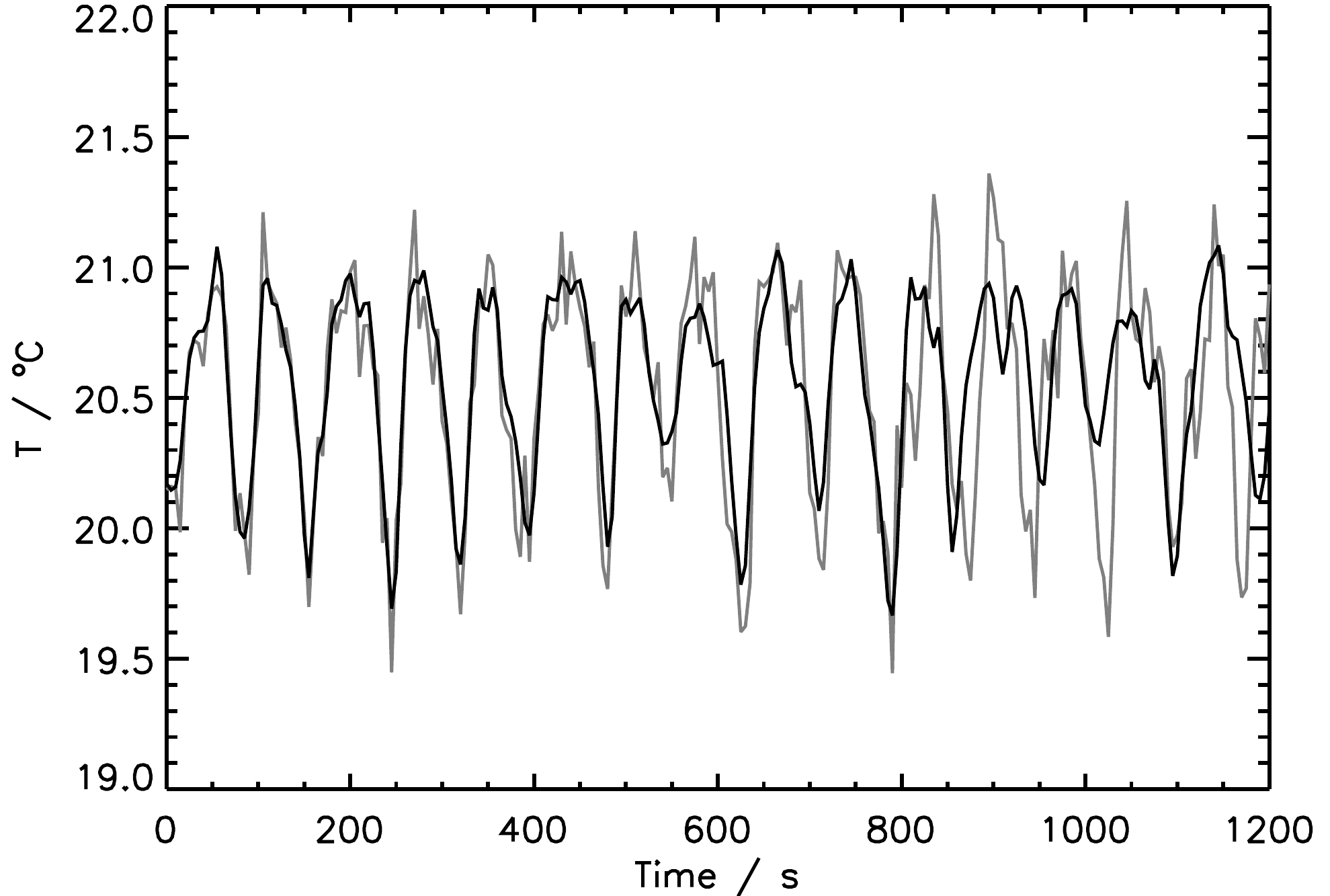}\label{fig:eg-T-tseries-tseries}}\\
  \subfloat[Temperature residual error (black line) and corresponding $\pm\sigma$ and $\pm 2\sigma$ observational error (dotted lines).]{\includegraphics[viewport=25 16 597 400,clip,width=0.8\textwidth]{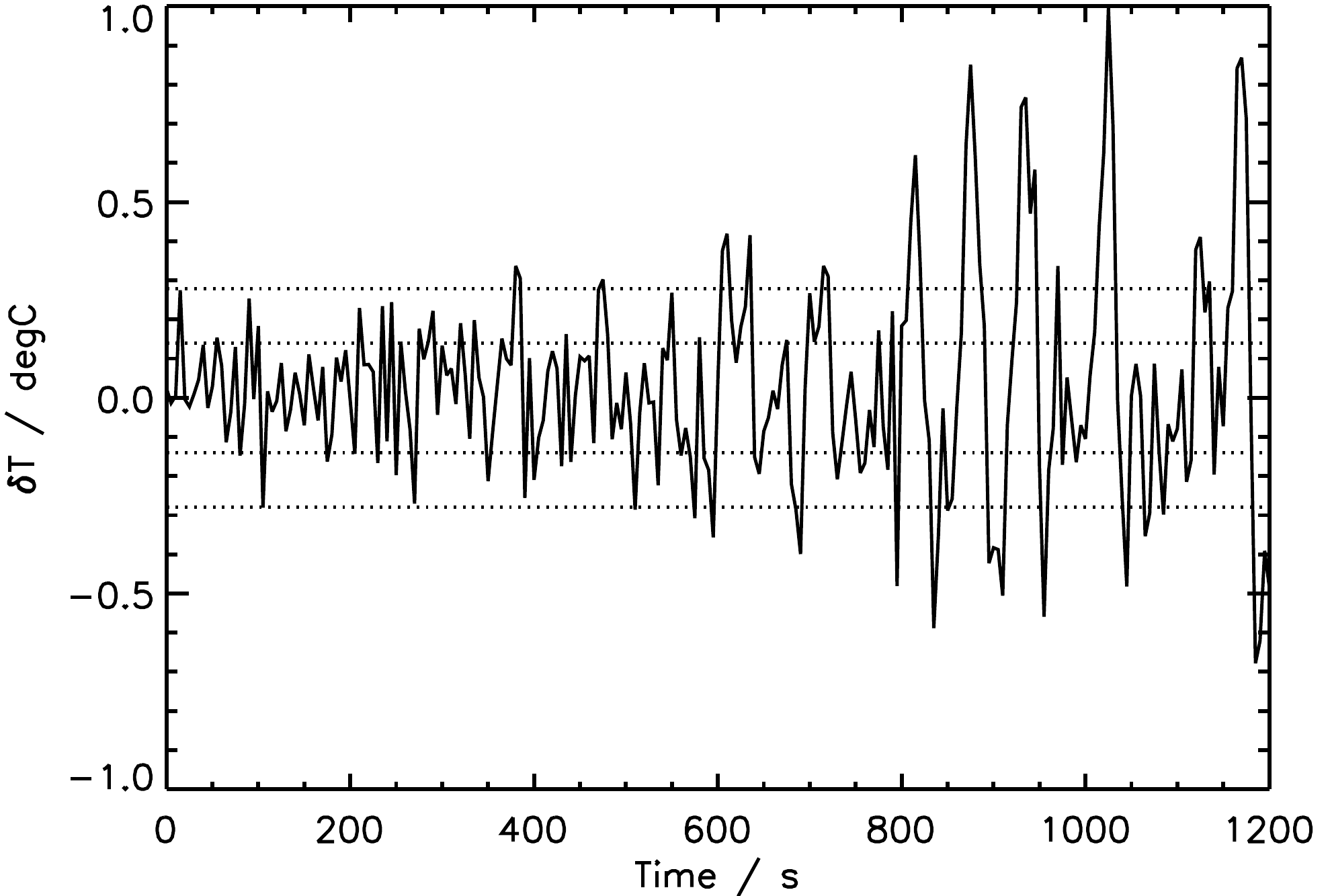}\label{fig:eg-T-tseries-residual}} 
  \caption{\small Example annulus candidate trajectory showing the temperature time series and residual error at a single grid point: $R=4.81$\,cm, $\theta=6.19$\,rad, $z=5.61$\,cm. This particular candidate trajectory begins a distance $\delta=0.1$ from the truth in the scaled space (Eq.~\ref{eq:delta}).}
  \label{fig:eg-T-tseries}
\end{figure}

We generated eight sets of initial candidate states using the method in Sect.~\ref{sec:perturb}, seven sets containing 64 candidates and one containing 256 (in order to have one with better statistics). While each set may be small compared with $N$, because the perturbation directions are random and initial conditions almost surely lie off the attractor, we would simply expect a smoothing of the distribution of candidate shadowing times if the number of candidates were increased further. Much longer or shorter shadowing times might occur in the special case of a perturbed initial condition falling on the attractor, but this occurs with probability zero. Within each set the distance from truth of each candidate was the same in the scaled space. We used $\delta=10^{-4}$, $10^{-3}$, $10^{-2}$,  $e.10^{-2}$, $e.10^{-1}$, $1$, and $e$ as the initial distances from truth in the scaled space for the 64-candidate sets, and $\delta=0.1$ for the set with 256 candidates\footnote{Multiples of $e$ were used to ensure approximately regular spacing of the perturbation sizes on a logarithmic scale.}. Using $\sigma=0.1$ the expected distance between observations and truth is $0.1\sqrt{24192}=15.6$ (Sect.~\ref{sec:iota-test}). Hence the largest distance between candidate state and truth ($\delta=e$) corresponds to about 15--20\% of the observational error, $\delta=0.1$ corresponds to about 0.6\%, and $\delta=10^{-4}$ corresponds to about 0.0005\% of the observational error. The candidate trajectories were integrated by MORALS for 1500\,s of simulated time, recording the state every 5\,s.

To calculate the appropriate $p$-values to use when measuring those candidates' shadowing times we used Eq.~(\ref{eq:no-type1}) with either $E=64$ or $E=256$ and $n=301$ (maximum trajectory length of 1500\,s, verifying every 5\,s, plus a verification at $t=0$). So for $R<1$ (fewer than one candidate suffering a Type I error) we found that $p<2.6\times 10^{-5}$ when $E=64$ and $p<6.5\times 10^{-6}$ when $E=256$, so we used $p=10^{-5}$ for the 64-candidate sets and $p=10^{-6}$ for the 256-candidate set.

Figure~\ref{fig:eg-T-tseries-tseries} shows one grid point in a typical MORALS candidate trajectory and corresponding (temperature) observations in the PMS using the setup described in Sect.~\ref{sec:model}. The residual error between the trajectory and observations is shown in Fig.~\ref{fig:eg-T-tseries-residual}, and the trajectory in Fig.~\ref{fig:eg-T-tseries-tseries} diverges from the observations at this particular grid point around 800\,s, corresponding to the time the residual in Fig.~\ref{fig:eg-T-tseries-residual} grows, on average, to be larger than the $\pm 2\sigma$ error bars.

\begin{figure}[tbp]
  \centering
  \includegraphics[viewport=2 2 597 
430,clip,width=0.7\textwidth]{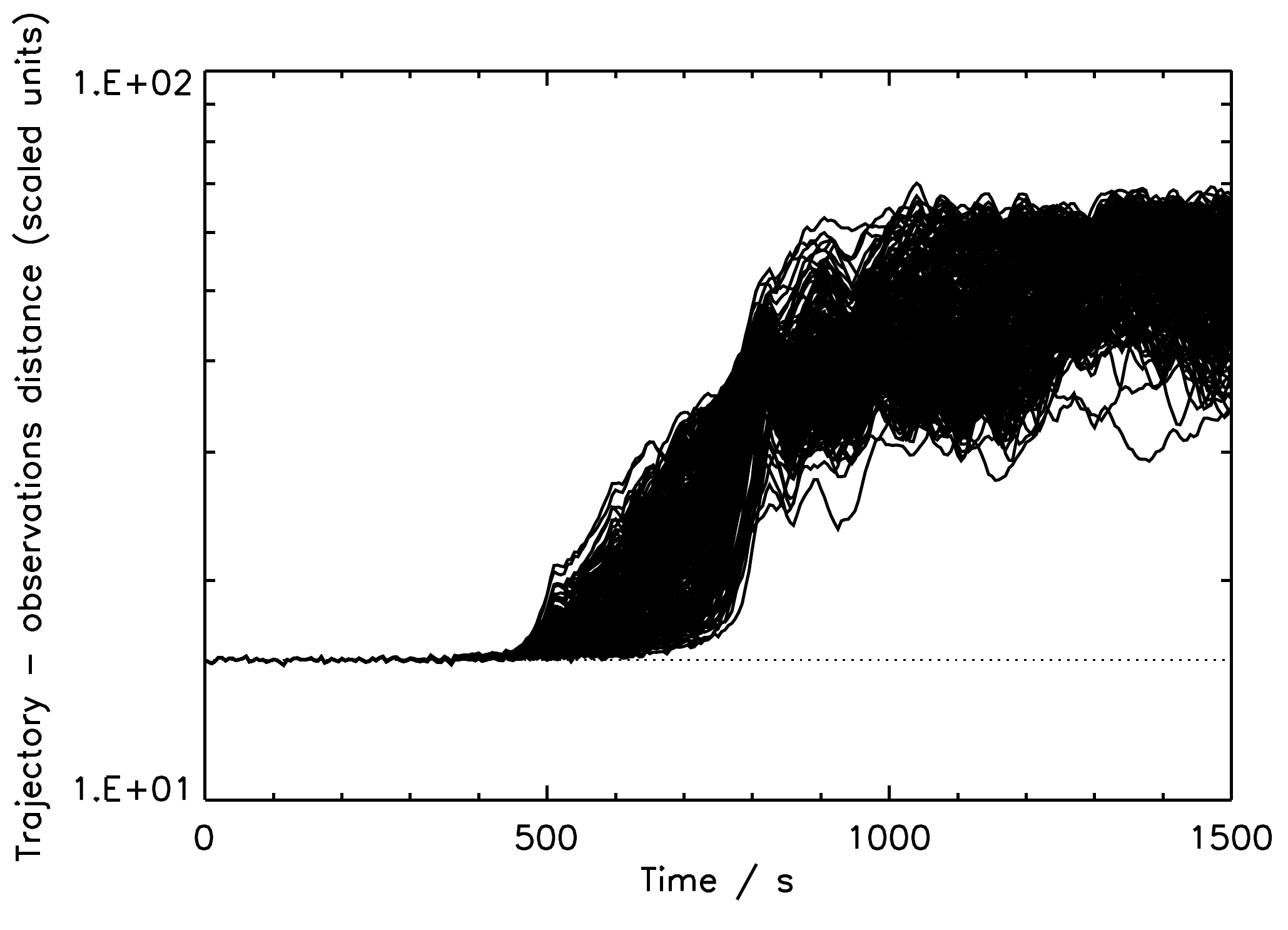}
  \caption{\small Distance between 256 candidate trajectories and one realisation of the observational noise (using Eq.~\ref{eq:fc-obs-dist}) for candidates begun a distance $\delta=0.1$ from truth in the scaled state space introduced in Eq.~(\ref{eq:scaled-space}). The vertical scale shows the Euclidean distance between candidate trajectory and observations in the scaled state space, and is logarithmic. The dotted line is at $D=\sigma\sqrt{N}$. This distribution of error growth over the candidates is robust with respect to different realisations of the observational noise (four were tried in total, with similar results).}
  \label{fig:pert2-dist}
\end{figure}

We measured the distance between the trajectories and observations as a function of time:
\begin{linenomath*}\begin{equation}
  D(t)=\Vert\{f^t({\bf x}[0])-{\bf s}[t]\}\circ{\bf r}^{-1}\Vert
  \label{eq:fc-obs-dist}
\end{equation}\end{linenomath*}
Figure~\ref{fig:pert2-dist} shows this distance for each trajectory in the 256-candidate set (an initial distance from truth of size $\delta=0.1$ in the scaled space). By construction, in the scaled space the expected initial distance between each candidate and the observations is $\sigma\sqrt{N}\sim 15.6$.

In this figure we see a rapid growth of the distance between candidate trajectories and observations from the expected initial observational error to the size of the attractor over a period of $\sim$300\,s between $t=500$\,s and 800\,s. Intuitively, we require our method to show that a candidate shadows observations until the distance between candidate and observations diverges from from the expected initial distance $\sigma\sqrt{N}$.

\begin{figure}[tbp]
  \centering
  \includegraphics[viewport=2 2 597 400,clip,width=0.7\textwidth]{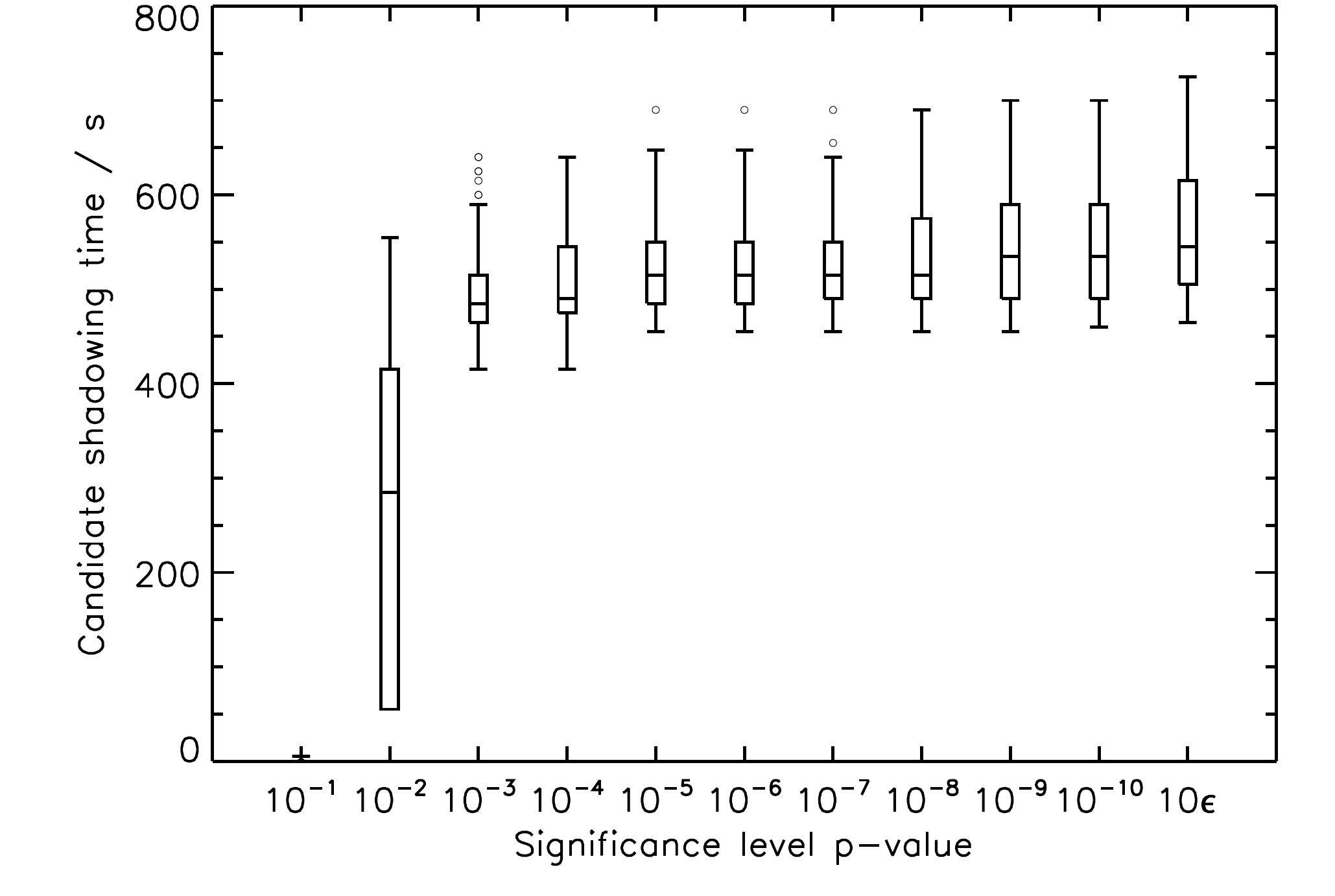}
  \caption{\small Distributions of candidate shadowing times measured using the ``state'' method for the 256 candidate trajectories in Fig.~\ref{fig:pert2-dist}  with $\delta=0.1$. Distributions for eleven different significance levels ($p$-values) are shown, with the $p$-value on the $x$-axis. Each box-whisker plot shows the distribution of candidate shadowing times at one significance level. The box represents the interquartile range, the middle line is the median, and the two whiskers represent the maximum/minimum or the 75\%/25\% values +/- 1.5 times the interquartile range, whichever is closest to the 75\%/25\% value. Outliers are shown with circles.}
  \label{fig:iota-shad-confs}
\end{figure}

Using the ``state'' method described in Sect.~\ref{sec:iota-test}, we measured the candidate shadowing times for each candidate in the eight sets, using a range of eleven significance levels in each case from $p=0.1$ up to the maximum significance permitted by floating point precision ($p=10\epsilon\approx 4.4\times 10^{-15}$). Figure~\ref{fig:iota-shad-confs} shows the distribution of candidate shadowing times for the trajectories in the 256-candidate set ($\delta=0.1$) for each of these significance levels.

Although above we calculated the ``appropriate'' significance level to use in this context, it is important to determine how sensitively the results depend on this quantity. The figure shows that our candidate shadowing times are generally not very sensitive to the significance level. Once the candidate trajectory diverges from the observations, it does so rapidly, reflecting the trajectory's rapid divergence from the observations as shown in Fig.~\ref{fig:pert2-dist}. As the residual error distribution diverges from the observational error distribution, the test rejects the null hypothesis even at the highest significance levels, and the robustness of our method in this respect is very encouraging. Below $p=0.001$ the candidate shadowing times are significantly smaller, which is expected from the analysis of the ``null'' trajectories above.

At $p=10^{-6}$, the ``appropriate'' significance level for this candidate set, the candidate shadowing times are in the range 450--700\,s. This concurs with a visual analysis of the trajectories in Fig.~\ref{fig:eg-T-tseries} and the visual estimate of about 500\,s from Fig.~\ref{fig:pert2-dist}. 

To determine whether the shadowing times are sensitive to position on the model attractor, we compared the candidate shadowing time distributions against two candidate trajectory sets started from other points in the state space. To start candidate trajectories from other points in the state space, the spinup time was changed from 1500\,s to 1525\,s and 1550\,s. At this point in $(\Omega,\Delta T)$ parameter space the timescale of the fluid's major oscillation (which determines the extent of the model attractor) is about 75\,s (Fig.~\ref{fig:eg-T-tseries-tseries}), so trajectories started from these other two spinup times are well-spaced about the model attractor. At each position we ran two 64-candidate sets with $\delta=10^{-3}$ and $e.10^{-1}$. The distributions of candidate shadowing times for $p=10^{-6}$ were very similar, with extrema and medians varying by only 5-10\%. Hence the distribution of candidate shadowing times is not generally sensitive to position in state space, at least for the points tested.

Now we have established that our method measures reasonable candidate shadowing times, we consider what the expected candidate shadowing time should be given its initial distance from truth. Conversely, how close to the truth must a candidate trajectory begin at to be expected to shadow for a particular time? With our perturbation experiments we can approach this question in a controlled fashion. 

\begin{figure}[tb]
  \centering
  \includegraphics[viewport=2 2 597 400,clip,width=0.7\textwidth]{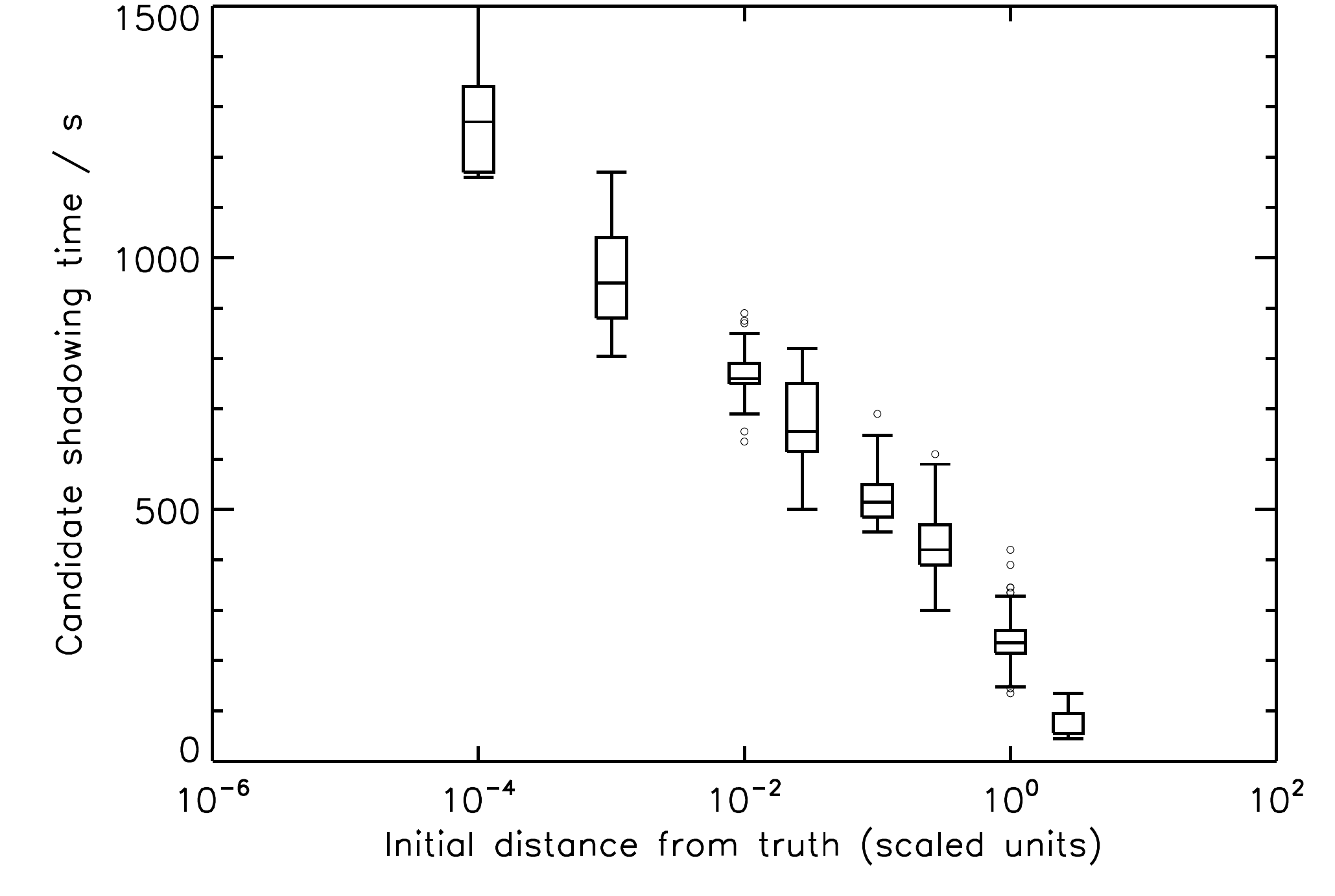}
  \caption{\small Distributions of candidate shadowing times for eight initial distances from truth $\delta$, using significance level $p=10^{-5}$ for all but $\delta=0.1$, which uses $p=10^{-6}$. The boxes are constructed in the same way as in Fig.~\ref{fig:iota-shad-confs}.}
  \label{fig:iota-shad-deltas}
\end{figure}

Using $p=10^{-5}$ for the 64-candidate sets, and $p=10^{-6}$ for the 256-candidate set, we plot in Fig.~\ref{fig:iota-shad-deltas} the distribution of candidate shadowing times for the eight different initial distances from truth $\delta$. There is clearly a logarithmic relationship between these candidate shadowing times and the initial distance from truth. An approximate empirical best fit line to the data is 
\begin{linenomath*}\begin{equation}
  \tau_{S}\approx 250\left(1-\log_{10}\delta\right)
  \label{eq:tau-vs-delta}
\end{equation}\end{linenomath*}
where $\tau_{S}$ is the candidate shadowing time, measured in seconds, and $\delta$ is the initial distance from truth (Eq.~\ref{eq:delta}). Conversely, we predict that a candidate will have an expected shadowing time of $\tau_S$ when the initial distance from truth in the scaled space is approximately
\begin{linenomath*}\begin{equation}
  \delta\approx10^{1-\tau_{S}/250}
  \label{eq:D}
\end{equation}\end{linenomath*}
To test this prediction, we ran a final set of candidate trajectories which we expect to shadow for around $\tau_{S}=2000$\,s. From Eq.~(\ref{eq:D}) this corresponds to $\delta\approx 10^{-7}$. We started the final 64-candidate set from the same point in state space as the previous eight sets, with an initial distance from truth of $\delta=10^{-7}$, and found the distribution of candidate shadowing times to be close to the predicted 2000\,s:

\begin{center}
\begin{tabular}{llllll}
\toprule
   Percentile (\%) & 0 & 25 & 50 & 75 & 100\\
   $\tau_{S}$ (s) & 1925 & 2000 & 2090 & 2130 & 2405\\
   \bottomrule
\end{tabular}
\end{center}

Note that the maximum shadowing time $\tau_{S_t}$ over all possible candidates is not what is being measured here. Candidates beginning on the attractor are special cases that the dynamics allows for but will almost never occur with our perturbation method. We loosely use the term ``on the attractor'' here to include points on the attractor and on the relevant stable manifold. In these cases a candidate further away from the truth may have a longer shadowing time, but our perturbation method generates an initial condition on the attractor with probability zero. Hence we can only conclude that the expected candidate shadowing time for a trajectory beginning a given distance from truth varies logarithmically with that distance; the maximum shadowing time may be much longer. 

Our aim was to demonstrate that the method behaves sensibly, and it does so: the statistical distribution of shadowing times decreases as initial distance from the truth increases. The functional form of this relationship is close to logarithmic, however, which is perhaps a little surprising. \citet{1999SmithA} showed that, even for low-dimensional systems like the Lorenz equations, the error $q$-pling time $\tau_q$ behaves such that the error-quadrupling time $\tau_4$ is generally much less than twice the error-doubling time $\tau_2$, indicating explosive and episodic uncertainty growth (their Fig.~8). By contrast, exponential divergence of two trajectories implies $\tau_{q^2}\approx 2\tau_q$. So one would not necessarily expect an logarithmic dependence of the shadowing time on initial distance from truth, as that would be equivalent to $\tau_{q^2}\approx 2\tau_q$. This is what we have observed, however, in Fig.~\ref{fig:iota-shad-deltas}. Our initial perturbations are possibly too large for this phenomenon to be relevant; \citet{1999SmithA} use infinitesimal perturbations in their analysis. Furthermore, they state that $\tau_{q^2}\approx 2\tau_q$ may hold once $q$ is large enough. $q=32$ is too small for the systems they use, but our experiments represent $q$ as high as $\approx 400$, which may be sufficient.

\subsection{Sanity check using the simple distance-based metric}

In Sect.~\ref{sec:phi-test} we introduced a simple distance-based metric to determine the time $\tau_D$ that a candidate trajectory remains within $m\sigma\sqrt{N}$ of the observations in the scaled state space (Eqs~\ref{eq:distance} and \ref{eq:phi-criterion}). This estimates the time at which the candidate trajectory diverges from the observations. 

\begin{figure}[tb]
  \centering
  \includegraphics[viewport=2 2 852 568,clip,width=0.7\textwidth]{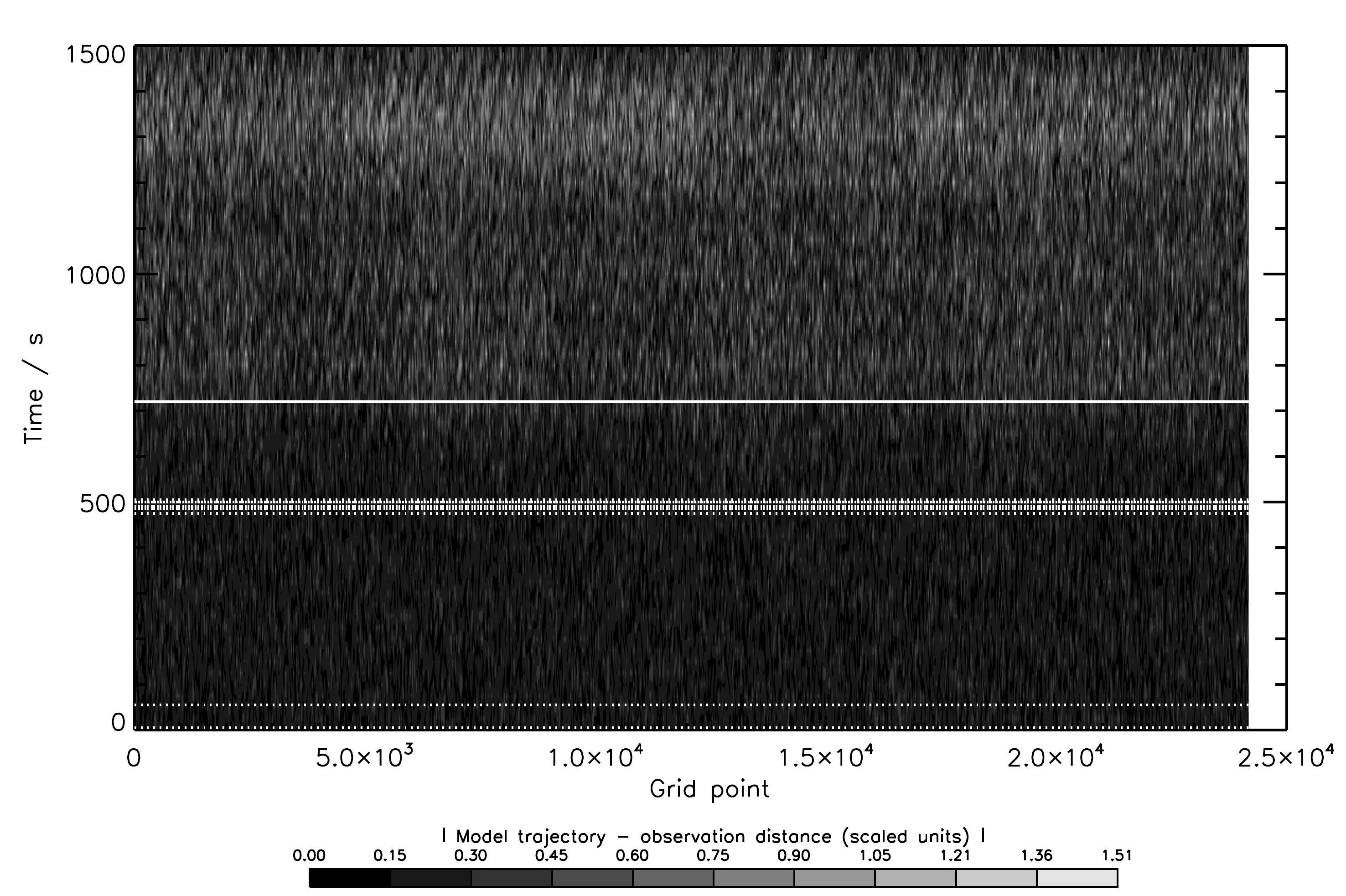}
  \caption{\small Contour plot of the unsigned distance $|{\bf e}[t]\circ{\bf r}^{-1}|$ between candidate trajectory and observations for each grid point in a candidate with $\delta=0.1$ (same run as Fig.~\ref{fig:eg-T-tseries}), for each point in the trajectory. A time series for a single grid point is a single vertical line in this diagram, and for a particular $t$ all the model grid points are in a horizontal line (the order of these grid points along the $x$-axis is unimportant). The horizontal solid line is at $t=\tau_D$ for this candidate (using $m=2$), and the dotted lines show $\tau_S$ for the 11 different $p$-values in Fig.~\ref{fig:iota-shad-confs}. Note that to enhance the contrast, the contour intervals are irregularly spaced.}
  \label{fig:pert2_p0001_contour}
\end{figure}

The trajectory and residual in Fig.~\ref{fig:eg-T-tseries} show an example at a single model grid point, where the model trajectory diverges from the observations around $t=800$\,s, but the definition of $\tau_D$ should identify the time at which this occurs for the \textit{whole} state vector. In Fig.~\ref{fig:pert2_p0001_contour} we show a visualisation of the same candidate trajectory (of which Fig.~\ref{fig:eg-T-tseries} was a single grid point) that extends the visualisation of residual error at one grid point to the whole state vector. There is an abrupt transition from small distances (dark regions) to large distances (light regions) over the whole state vector, quite similar over most of the grid points, which is the time at which the candidate trajectory diverges from the observations. Below we list several percentiles of the probability distribution function for $\tau_D$ measured for each trajectory in the 256-candidate set. 

\begin{center}
\begin{tabular}{llllll}
\toprule
   Percentile (\%) & 0 & 25 & 50 & 75 & 100\\
   Shadowing time (s) & 675 & 735 & 760 & 775 & 980\\
   \bottomrule
\end{tabular}
\end{center}

In Fig.~\ref{fig:pert2_p0001_contour}, $\tau_D$ is identified by a solid horizontal line. There is general agreement between $\tau_D$ and the time at which the absolute distance between candidate trajectory and observations increases rapidly, i.e. from dark to light in the figure as we move upwards. $\tau_D$ corresponds well to this transition for all candidates for which such diagrams were produced, being within $\pm 50$\,s in each case. Hence $\tau_D$ estimates well the time at which a candidate trajectory diverges from observations. Furthermore, the time between diverging from the expected distance between observations and truth and reaching the saturation distance (distance between two points randomly chosen on the attractor) is relatively short compared with the length of a trajectory (Fig.~\ref{fig:pert2-dist}). So as long as $m$ is not so large as to make $m\sigma\sqrt{N}$ larger than the saturation distance, then $\tau_D$ will be quite insensitive to $m$.

The dotted lines show $\tau_S$ measured in the previous section for this particular candidate for each of the eleven $p$-values. In this case, as in the others, $\tau_S<\tau_D$, which demonstrates that the distribution of residual error loses consistency with the observational noise distribution some time before the trajectory and observations diverge visually. Overall, however, our simple distance metric confirms that our method for measuring shadowing times is a reasonable measurement of the time model trajectories diverge from observations.

\section{Discussion and conclusions}
\label{sec:conc}

We have presented a study of shadowing in the rotating annulus using the perfect model scenario. First we developed a version of the \citet{2010Smith} ``sequence'' method of calculating the shadowing time that is more suitable for use with high-dimensional systems. In the course of this analysis we found a way to avoid the practical sampling problems of applying the sequence method in higher dimensions.  Our ``state'' method measures the consistency of the residual error distribution and the observational error distribution, and requires consistency at each time up to $t$ in order to shadow for a time $t$.

We demonstrated our state method using rotating annulus simulations in a series of perturbation experiments. Generating sets of candidate states a fixed distance from the true state, we measured the candidate shadowing times with respect to a set of artificial observations. Because our perturbation method ensured that, with probability one, initial conditions would not lie on the model attractor or stable manifold, we expected (almost surely) the trajectories to diverge from the observations, and they did so. The shadowing times decreased as the trajectories' initial distances from truth increased, with a clear logarithmic relationship between the initial distance from truth and the distribution of candidate shadowing times. We used the relationship to predict how close to the truth the candidate would have to be for its expected shadowing time to be 2000\,s, and the results agreed well with the prediction. The candidate shadowing times (Fig.~\ref{fig:iota-shad-confs})  also corresponded well to the time trajectories remained close to observations measured using a simple distance metric (Fig.~\ref{fig:pert2_p0001_contour}), concurred with a visual analysis of the forecasts (Fig.~\ref{fig:eg-T-tseries}), and appeared to be similarly distributed at other points in the model state space, all of which strengthen our conclusion that the method we have presented to measure shadowing times is reasonable.

By using the PMS we were able to measure shadowing times against the true state of the system using a simple setup. In particular, deliberately simplifying the problem in this way avoided complications introduced by the $\epsilon$-shadowing properties of the model, which are currently unknown. We expect the model's equations to be non-hyperbolic and so long $\epsilon$-shadowing trajectories probably don't exist, for various reasons \citep{1999Lai}. The larger the number of dimensions the more complications this is expected to introduce: for example, see \citet{2003Hayes} for an analysis in $O(100)$ dimensions, which is still a factor of $O(100)$ smaller than MORALS. To investigate these properties further, one could determine whether phenomena such as glitching \citep{1991SauerB,1994Dawson} or unstable-dimension variability \citep{1997Kostelich} might be important. \citet{1994Dawson} showed that if a finite-time local Lyapunov exponent fluctuates about zero then $\epsilon$-shadowing will fail quickly, and so quantifying this using the hyperbolicity exponent \citep{2002Sauer}, which is possible in principle although difficult in practice, would provide some insight. Another simple experiment would be to compare a ``clean'' model trajectory with one where noise on the order of the roundoff error is introduced at each timestep, which would give a lower bound on the $\epsilon$-shadowing times. This approach has precedent in \citet{1997Sauer}; see also \citet{1987Smith} for an example using numerical representations of 1--1 chaotic maps without round-off error. In any case, we expect the model's intrinsic structural error to be larger than these other sources of error, so in practice shadowability of the real experiment will almost certainly be limited by model error rather than anything else.

These results define a benchmark for further study of shadowing in this system. From Eq.~(\ref{eq:D}) we have a quantitative measure of whether a candidate shadowing time of, say, 400\,s is ``good'' or ``bad'', because it tells us how close to the true state we should expect a randomly orientated perturbation to be in order to expect that candidate shadowing time. In addition, in the PMS the candidate shadowing time tells us approximately how close to the truth our candidate was initially, assuming random perturbations, and hence how much we might expect to be able to improve it. Because of our perturbation method our candidate shadowing times will be lower bounds in the PMS --- a method that generates candidates close to or on the attractor (such as gradient descent) should yield longer candidate shadowing times. With an imperfect model or observational data, however, we would expect these shadowing times to be an upper bound given that then the model and system attractors will then be disjoint or almost disjoint.

It should be stressed that these experiments were performed with one particular set of model parameters. The particular candidate shadowing times obtained would therefore be expected to change for different points in the parameter space. In a non-chaotic flow regime, for example, it is conceivable that a candidate trajectory might shadow observations for all $t$. Nevertheless, in other chaotic regions of parameter space we predict that the candidate shadowing times should vary in a similar way as a function of initial distance from the truth.

The work we have presented in this paper falls into a larger body of work examining the shadowing properties of the rotating annulus system by applying the method of gradient descent \citep{2003Judd}. This method searches for shadowing trajectories by relaxing a sequence of observational states onto the model attractor. Gradient descent has been used in low- \citep[and elsewhere]{2003Judd}, intermediate- \citep{2004JuddB} and high- \citep{2008JuddA} dimensional models for the purposes of state estimation. We intend to combine our ``state'' method for measuring shadowing times with an algorithm for gradient descent and use this to determine whether gradient descent can produce candidates with long shadowing times in the rotating annulus system. In the perfect model scenario we will know how far away a state obtained via gradient descent is from the truth, and hence from Eq.~(\ref{eq:tau-vs-delta}) we will have an \textit{a\,priori} estimate of how long we might expect a candidate to shadow observations. This work will be a useful step towards determining whether the gradient descent method can produce model trajectories with long shadowing times in high-dimensional models of weather and climate. 

\section*{Acknowledgements}

We thank Leonard Smith for providing the original impetus for this work and for numerous insights during its duration. We thank him and Hailang Du for reading the manuscript fully more than once. We also thank Kevin Judd, Peter Read, and Thomas Stemler for useful conversations on a number of topics. We thank two anonymous reviewers whose comments have greatly helped us to improve the manuscript. The program used to compute Lyapunov exponents was written by Peter Read, and the TISEAN package is available from http://www.mpipks-dresden.mpg.de/$\sim$tisean/. RMBY and FN acknowledge financial support from the Grantham Research Institute on Climate Change and the Environment. RMBY also acknowledges financial support from NERC Studentship NER/S/A/2005/13667.

\appendix

\section{MORALS equations}
\label{app:morals-eq}

MORALS solves the fluid equations in cylindrical polars $(R,\phi,z)$ for a 3D velocity field ${\bf u}=(u,v,w)$ and temperature $T$ in a reference frame rotating at angular velocity $\Omega$ using a finite-difference integration scheme with leapfrog time-stepping. The pressure $\Pi$ is found diagnostically using a Poisson equation \citep[Sect.~5.1]{1976Farnell}. $T$ is relative to a reference temperature $T_{\rm R}=22$\,$^{\circ}$C, and $\Pi$ is relative to a reference pressure \mbox{$\Pi_0(R,z)=\frac{1}{2}\Omega^2 R^2+g(d-z)$}. The equations solved are the continuity, heat, and Navier-Stokes equations:
\begin{linenomath*}\begin{equation}
  \nabla\cdot{\bf u}=0
\end{equation}\end{linenomath*}
\begin{linenomath*}\begin{equation}
  \frac{\partial T}{\partial t}+{\bf u}\cdot\nabla T=\nabla\cdot(\kappa\nabla T)
\end{equation}\end{linenomath*}
\begin{linenomath*}\begin{equation}
  \frac{\partial {\bf u}}{\partial t}+{\bf u}\cdot\nabla{\bf u}+2{\bf \Omega}\times{\bf u}-(\Omega^2 R\hat{\bf R}-g\hat{\bf z})\frac{\Delta\rho}{\rho_0}+\nabla\Pi={\bf F}
\end{equation}\end{linenomath*}
Here $\bf F$ is the viscous term, but is more complicated than the standard $\nu\nabla^2{\bf u}$ because it allows spatial variations of the viscosity as well; see \citet[Eqs~2.1--2.3, 2.6]{1976Farnell} for the full expression. These equations are closed by an equation of state for density $\rho(T)=\rho_0(1+\rho_1T+\rho_2T^2)$ with $\Delta\rho=\rho-\rho_0$ and two ``constitutive'' relations for viscosity $\nu(T)=\nu_0(1+\nu_1T+\nu_2T^2)$ and thermal diffusivity $\kappa(T)=\kappa_0(1+\kappa_1T)$. On all boundaries ${\bf u}={\bf 0}$ and $\hat{\bf n}\cdot\nabla\Pi=0$, on the top and bottom boundaries $\hat{\bf n}\cdot\nabla T=0$, and on the inner and outer cylinders $T=T_a-T_R$ and $T=T_b-T_R$ respectively. 

\bibliography{012-013-LSE}
\bibliographystyle{agufull08-ed}

\end{document}